\documentclass[aps,prd,11pt, tightenlines, twoside, secnumarabic, superscriptaddress, showpacs, preprintnumbers, nofootinbib, notitlepage, onecolumn, fleqn, longbibliography, a4paper]{revtex4-1}
\pdfoutput=1
\usepackage[english]{babel}
\usepackage{amsmath,amssymb,amsfonts,bm,bbm,slashed}
\usepackage{graphicx}
\usepackage[sort&compress]{natbib}
\usepackage{xcolor}
\usepackage[normalem]{ulem}
\usepackage{hyperref}
\usepackage{cleveref}
\definecolor{red}{rgb}{1.0, 0, 0}
\usepackage{enumerate}
\usepackage{epsfig, subfigure}
\usepackage{setspace}
\usepackage{booktabs, tabularx}
\usepackage{units}
\usepackage[utf8]{inputenc}
\usepackage{lineno}
\usepackage{feynmp-auto}               

\hypersetup{
        pdfnewwindow=true,   
        colorlinks=true,     
        linkcolor=blue,      
        citecolor=blue,      
        filecolor=blue,      
        urlcolor=blue        
}

\allowdisplaybreaks

\bibpunct{[}{]}{,}{n}{}{,}

\newcommand{\ev}[1]{\ensuremath{\left\langle #1 %
                \right\rangle}} 

\renewcommand{\vec}[1]{{\mathbf{#1}}}

\newcommand{\vev}[1]{\ev{#1}}

\newcommand{\be}[2]{\left.  #1 \right|_{#2}}              

{

\begin{document}
        
\title{Dynamic Freeze-In: Impact of Thermal Masses and Cosmological Phase
           Transitions on Dark Matter Production}

\author{Michael J.\ Baker}
\email{baker@physik.uzh.ch}
\affiliation{Physik-Institut, Universit\"at Z\"urich, 8057 Z\"urich, Switzerland}

\author{Moritz Breitbach}
\email{mbreitba@students.uni-mainz.de}
\affiliation{PRISMA Cluster of Excellence \& Mainz Institute for Theoretical Physics,
        Johannes Gutenberg University, Staudingerweg 7, 55099 Mainz, Germany}

\author{Joachim Kopp}
\email{jkopp@uni-mainz.de}
\affiliation{PRISMA Cluster of Excellence \& Mainz Institute for Theoretical Physics,
        Johannes Gutenberg University, Staudingerweg 7, 55099 Mainz, Germany}

\author{Lukas Mittnacht}
\email{lmittnac@students.uni-mainz.de}
\affiliation{PRISMA Cluster of Excellence \& Mainz Institute for Theoretical Physics,
        Johannes Gutenberg University, Staudingerweg 7, 55099 Mainz, Germany}

\date{\today}
\preprint{MITP/17-098}
\preprint{ZU-TH-37-17}

\begin{abstract}
  The cosmological abundance of dark matter can be significantly influenced by
  the temperature dependence of particle masses and
  vacuum expectation values.  We illustrate this point in three simple
  freeze-in models. The first one, which we call kinematically induced
  freeze-in, is based on the observation that the effective mass of a scalar
  temporarily becomes very small as the scalar potential undergoes a 
  second order phase transition. This opens dark matter production channels that are
  otherwise forbidden.  The second model we consider, dubbed vev-induced
  freeze-in, is a fermionic Higgs portal scenario.  Its scalar sector is
  augmented compared to the Standard Model by an additional scalar singlet, $S$,
  which couples to dark matter and temporarily acquires a vacuum expectation
  value (a two-step phase transition or ``vev flip-flop'').  While $\ev{S} \neq
  0$, the modified coupling structure in the scalar sector implies that dark
  matter production is significantly enhanced compared to the $\ev{S} = 0$
  phases realised at very early times and again today.  The third model, which
  we call mixing-induced freeze-in, is similar in spirit, but here it is the
  mixing of dark sector fermions, induced by non-zero $\ev{S}$, that
  temporarily boosts the dark matter production rate.  For all three scenarios,
  we carefully dissect the evolution of the dark sector in the early Universe.
  We compute the DM relic abundance as a function of the model parameters,
  emphasising the importance of thermal corrections and the proper treatment of
  phase transitions in the calculation.
\end{abstract}

\maketitle 

\section{Introduction}
\label{sec:intro}

``In order for the light to shine so brightly, the darkness must be present.''
This quote, attributed to Sir Francis Bacon, subsumes much of modern day
cosmology.  The Universe as we know it, with its abundance of bright galaxies,
could not have formed without the presence of large amounts of dark matter
(DM). DM drives the formation of structure; the gravitational collapse
of primordial density fluctuations leads to dense objects like galaxy clusters,
galaxies, and stars, with at least one of the latter harbouring life.  Even
though one of the lifeforms likes to describe itself as intelligent, it is
still very much in the dark about dark matter, its origin and its nature.

For a long time, the best-motivated scenario to understand DM has been
the Weakly Interacting Massive Particle (WIMP) scenario: DM particles 
are hypothesised to be heavy ($m_\text{DM} \gtrsim 100$~GeV) and to have weak, but
non-negligible interactions with Standard Model (SM) particles.  In the
very early Universe, these interactions keep the DM and SM sectors
in kinetic and chemical equilibrium, until eventually Hubble expansion dilutes
the DM density to the extent that DM annihilation into
SM particles ceases. This \emph{freeze-out} typically occurs at
temperatures $T$ where $x \equiv m_\text{DM} / T \simeq 25$~\cite{Kolb:1990}.

While freeze-out is arguably still the leading scenario for explaining
the DM abundance in the Universe, the lack of experimental evidence for
DM particles, in spite of a vigorous, multi-pronged search 
program~\cite{Akerib:2016vxi, Tan:2016zwf, Ackermann:2015zua,
Accardo:2014lma, Madhavacheril:2013cna, Aaboud:2016tnv,
CMS:2016pod},
motivates the study of alternative mechanisms~\cite{Feng:2008ya,
Peccei:1977hh, Duffy:2009ig, Adhikari:2016bei, Hochberg:2014dra, Hochberg:2014kqa, 
Kuflik:2015isi, Bird:2016dcv, Giudice:2000ex, McDonald:2001vt, 
Hall:2009bx, Blennow:2013jba}.
Freeze-in models assume that the initial DM abundance after
inflation was zero and that DM particles couple
so weakly to the particles in the thermal bath that they never reach 
thermal equilibrium~\cite{Giudice:2000ex, McDonald:2001vt, Hall:2009bx, Blennow:2013jba, Bernal:2017kxu}. 
Consequently,
DM annihilation does not determine the relic abundance.
Instead, the observed abundance is the result of processes with 
DM in the final state, typically at $x \simeq 1$--$5$.  
The small coupling between DM and SM particles also implies a significantly 
greater challenge for all experimental probes of the nature of DM.

In this paper, we consider the impact of finite temperature
effects on the DM abundance in freeze-in models. Such effects are manifold:
first, particle masses receive corrections from thermal loops, implying
that the kinematics of DM production is in general $T$-dependent.
Certain production channels may be open during some epochs of cosmological
history, but kinematically closed during others.  Closely related
to this effect is the $T$ dependence of the effective scalar potential
$V^\text{eff}$, which implies that not only scalar masses (the second
derivatives of $V^\text{eff}$), but also
their vacuum expectation values (vevs) change with time in the early Universe.
These vevs, in turn, will affect gauge boson and fermion masses (of course, gauge
boson and fermion masses also receive direct corrections from self-energy diagrams
evaluated at $T \neq 0$, although the contribution to fermion masses 
will be unimportant in the scenarios we discuss). 
The most interesting case is where the scalar potential develops
several disjoint minima and transitions from one to another in a
phase transition.  The phase transition, for which the scalar vev is
an order parameter, can be first order, second order, or a mere cross-over.
The latter (and perhaps least interesting case) is realised in the 
SM~\cite{Arnold:1992rz, Kajantie:1996qd, Kajantie:1996mn, Csikor:1998eu, Rummukainen:1998as}.

In standard freeze-out scenarios, thermal effects are usually negligible since
they are small at $x \gg 1$, i.e., at temperatures much lower than the masses
of the involved particles.  For freeze-in at $x \simeq 1$--$5$, on the other
hand, they can be large and have a decisive impact on the DM abundance.
Demonstrating this with several examples is the main topic of this paper.

To focus on the important effects and avoid unnecessary complications, we
consider a toy model consisting of the standard model, a new gauge singlet
scalar, and one or two gauge singlet dark sector fermions.  Since we focus on
freeze-in, we imagine some couplings to be $\ll 1$.  As discussed below, these
small couplings will often be technically natural (i.e., protected from large
radiative corrections).  In other cases, we assume a small coupling to simplify
the analysis, but in these cases we do not expect a larger coupling to spoil
the overall picture.  The scenarios we consider could be motivated from a wide
range of UV theories, including SUSY and extra dimensional
scenarios~\cite{Hall:2009bx}.  Scenarios with gauge singlets at the weak scale
are notoriously difficult to test at colliders and will not be ruled out at the
LHC.  The scenarios discussed in \cref{sec:vev,sec:mix} will be best probed
through their $\mathcal{O}(1)$ Higgs portal couplings, but the full parameter
space will not be probed until a 100~TeV collider has collected 3~ab$^{-1}$ of
integrated luminosity~\cite{Curtin:2014jma}.  The scenario discussed in
\cref{sec:kin} is even harder to  exclude, due to the small Higgs portal
coupling.

The impact of thermal effects in the early Universe on the DM abundance has
been previously discussed for instance in \cite{Baker:2016xzo, J.Baker:2018eaq}, where
it was argued that DM could be temporarily unstable in the early Universe,
so that its abundance would be controlled by its decay rates and by
the temperature of the phase transition that stabilises it. Similar thermal effects 
have also been considered in, e.g.,~\cite{Rychkov:2007uq, Strumia:2010aa}.

The outline of the paper is as follows.  In \cref{sec:kin} we discuss a
scenario, dubbed ``kinematically induced freeze-in'', in which the
kinematics of DM production is controlled by the $T$-dependent masses of a new
scalar $S$ and a new dark sector fermion $\psi$. 
DM particles $\chi$ freeze in through their couplings to $S$ and 
$\psi$, but for most of cosmological history DM production via
$\psi \to S \chi$ is kinematically forbidden. However, as $S$ transitions from
a phase with $\ev{S} = 0$ to a phase with $\ev{S} \neq 0$, its mass drops close
to zero and DM production becomes kinematically allowed for a short period of
time.  We emphasise that, in a more economical version of this model, $S$ could
be replaced by the SM Higgs itself.  In \cref{sec:vev} we
consider an alternative freeze-in model --- a variant of the fermionic Higgs
portal scenario --- in which the DM production rate in the dominant channels is
proportional to $\ev{S}$. We call this scenario ``vev-induced freeze-in''. 
There is a large region of parameter space where its
scalar potential undergoes a two-step phase transition (``vev flip-flop''),
i.e., the Universe starts out in a $\ev{S} = 0$ phase, followed by an epoch
with $\ev{S} \neq 0$. The electroweak phase transition ends this epoch and
reverts the Universe to $\ev{S} = 0$ (but a non-zero vev for the SM Higgs).  It
is thus the two phase transitions that control the DM abundance today.  In a
third model, which will be the topic of \cref{sec:mix}, $\ev{S}$
controls mixing between the DM particle and a second new fermion.  This mixing,
in turn, opens up production channels that are otherwise inaccessible, thus
boosting freeze-in production.  Hence we call this scenario ``mixing induced
freeze-in''. We summarise and conclude in \cref{sec:conclusions}.

\section{Kinematically Induced Freeze-In:
         Temperature-Dependent Masses and Thresholds}
\label{sec:kin}

In thermal quantum field theory, particle masses can receive
temperature-dependent corrections from self-energy diagrams and thus become
functions of $T$ themselves.  For instance, the SM Higgs mass is $m_h(T=0) =
125\,\text{GeV}$ today, but was much larger in the very early Universe and
close to zero around the time of the electroweak cross-over at $T_\text{EW}
\simeq 165\,\text{GeV}$.  In this section we discuss a scenario where the
kinematics of the DM freeze-in rate are controlled by the mass of a new real
scalar $S$.

\subsection{Toy Model}
\label{sec:kin:model}

We consider a simple toy model --- a variant of the fermionic Higgs portal
scenario --- whose particle content is given in \cref{tab:kin-particles}.
Besides the real scalar $S$ and the Dirac fermion $\chi$, which is the DM candidate,
we introduce a second new Dirac fermion $\psi$.  All new particles are SM
singlets, and $\psi$ and $\chi$ are charged under a $\mathbb{Z}_2$ symmetry.
We remark already here that one could imagine a variation of the model in which
$S$ is replaced by the SM Higgs field itself.  The relevant terms in the
Lagrangian at dimension four are
\begin{align} 
  \mathcal{L} &\supset
      \frac{1}{2} (\partial_\mu S)(\partial^\mu S)
    + \bar{\psi}(i\slashed{\partial} - m_\psi)\psi 
    + \bar{\chi}(i\slashed{\partial} - m_\chi)\chi
                                            \nonumber\\[0.2cm]
  &\qquad
    + \big[ y_{\psi\chi} \, \bar\psi S \chi + h.c. \big]
    + y_\chi \, \bar\chi S \chi
    + y_\psi \, \bar\psi S \psi
    - V(H, S)
  \label{eq:kin:L}
\intertext{with}
  V(H, S) &=
   - \mu_H^2 H^\dag H
   + \lambda_{H4} \, (H^\dag H)^2
   - \frac{1}{2} \mu_S^2 S^2
   + \frac{\lambda_{S4}}{4!} S^4
                                            \nonumber\\[0.2cm]
  &\qquad
   + \frac{\lambda_{S3}}{3!} \, \mu_S S^3 
   + \lambda_{p3} \mu_S \, S (H^\dag H)
   + \frac{\lambda_{p4}}{2} \, S^2 (H^\dag H) \,.
  \label{eq:kin:V} 
\end{align}
The first line of \cref{eq:kin:L} contains the standard kinetic terms and the
fermion mass terms.  In the second line of \cref{eq:kin:L}, we identify the
Yukawa couplings between $S$, $\chi$ and $\psi$.
We assume $y_\chi$ and $y_{\psi\chi}$ to
be tiny to avoid full thermalisation of $\chi$ and thus allow for DM production
via freeze-in rather than freeze-out. The smallness of $y_\chi$ and
$y_{\psi\chi}$ could be motivated, for instance, by extra-dimensional scenarios
where $\chi$ could be localised far away from $\psi$ and $S$ along the fifth
dimension. The coupling $y_\psi$ on the other hand, is assumed to be sizeable so that
$\psi$ and $S$ remain in thermal equilibrium until $T \ll m_\psi,\, m_S$.
Alternatively, one can also introduce an extra particle -- for instance a second
new scalar $S'$ with negligible couplings to the DM particle $\chi$, but
appreciable couplings to $\psi$ and to the SM sector -- to achieve the same goal. In fact, in
the numerical results shown below we will assume this second possibility because
it simplifies the dynamics of the temperature-dependent effective scalar potential
and opens up larger regions of parameter space than the vanilla scenario from
\cref{eq:kin:L}.

The first line of the scalar potential $V(H, S)$ in \cref{eq:kin:V} contains the mass
terms and quartic couplings for $S$ and $H$. We assume $\mu_H^2\,, \mu_S^2 > 0$, so
that not only $H$, but also $S$, may obtain a vev at tree level.  The second line
of \cref{eq:kin:V} consists of a cubic coupling for $S$ proportional to $\lambda_{S3}$ 
and of the cubic ($\propto \lambda_{p3}$) and quartic ($\propto \lambda_{p4}$)
Higgs portal couplings.  We will assume that $\lambda_{S3}$, $\lambda_{p3}$,
and $\lambda_{p4}$ are $\ll 1$. For $\lambda_{S3}$ this is just a simplifying
assumption that could be relaxed at the expense of unnecessarily complicating
our analysis.  The smallness of $\lambda_{p3}$ and $\lambda_{p4}$ could again
be motivated in extra-dimensional scenarios by localising $S$ and $H$ far from
each other along the fifth dimension.  We hypothesise, however, that
$\lambda_{p4}$ is still large enough to keep $S$ in
thermal contact with the SM particles at temperatures $T \simeq m_S$, when DM
freeze-in happens. We find that these conditions are satisfied for
$\lambda_{p4} \sim 10^{-3}$.

\begin{table}
  \centering
  \begin{minipage}{10cm}
    \begin{ruledtabular}
    \begin{tabular}{cccc}
      Field  &      Spin     & $\mathbb{Z}_2$ & mass scale \\
      \hline
      $S$    &        0      &      $+1$      & $m_S(T=0)    \simeq 5\,\text{GeV}$ \\
      $\chi$ & $\frac{1}{2}$ &      $-1$      & $m_\chi \simeq 50\,\text{GeV}$ \\
      $\psi$ & $\frac{1}{2}$ &      $-1$      & $m_\psi \simeq 50\,\text{GeV}$ \\
      \bottomrule
    \end{tabular} 
    \end{ruledtabular}
  \end{minipage}
  \caption{The new particles we introduce in \cref{sec:kin} with their respective charges 
    and mass scales.  All new particles are SM gauge singlets.}
  \label{tab:kin-particles}
\end{table}

Where necessary, we will decompose $H$ into its components according to $H =
\big(G^+, (h + i G^0\big)/\sqrt{2})$, where $h$ is the neutral CP even SM-like Higgs
boson and $G^\pm$, $G^0$ are the Goldstone modes. Moreover, we will often use
the definitions $v_S \equiv \ev{S}$ and $v_H \equiv \ev{h}$ for the vacuum
expectation values of $S$ and $h$.

\begin{figure}
  \centering
  \includegraphics[width=\textwidth]{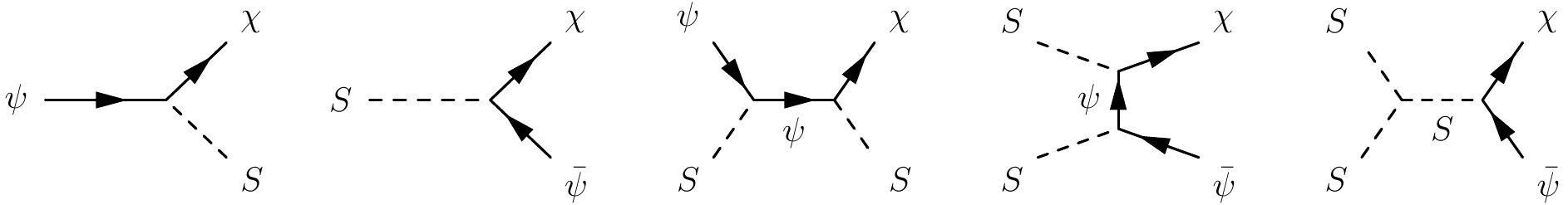}
  \caption{Dark matter freeze-in reactions in the toy model of
    kinematically induced freeze-in, see \cref{eq:kin:L}.
    For $S \psi \to S \chi$ and $S S \to \chi \bar\psi$, we do not
    show the diagrams with crossed $S$ lines.}
  \label{fig:kin:diagrams}
\end{figure}

Freeze-in of $\chi$ can proceed through the decays $\psi \to S \chi$ or $S \to \chi
\bar\psi$, (depending on the relative magnitude of the $\chi$, $S$, and $\psi$
masses), and through the $2 \to 2$ reactions $\psi S \to \chi S$ and $S S
\to \chi \bar\psi$.  The Feynman diagrams for these four processes are depicted in
\cref{fig:kin:diagrams}.  We will focus on masses such that $m_\psi + m_\chi >
m_S(T=0)$ and $m_\psi - m_\chi < m_S(T=0)$.  This implies that the decays $\psi
\to S \chi$ and $S \to \chi \bar\psi$ are kinematically forbidden today.  In
the very early Universe, however, $m_S$ receives large thermal corrections
$\propto T$, which can lift its value above $m_\psi + m_\chi$ and thus open up
the channel $S \to \chi \bar\psi$. Later, around the time when $S$ develops a
non-zero vev, $m_S(T)$ approaches zero and the channel $\psi \to S \chi$
becomes temporarily available.
The decay
rates and annihilation cross sections for the processes in
\cref{fig:kin:diagrams} are
\begin{align}
  \Gamma(\psi \to \chi S)
    &= \frac{y_{\psi\chi}^2}{16 \pi}
       \frac{(m_\chi + m_\psi)^2 - m_S^2(T)}{m_\psi^3}
				\nonumber\\
    &\hspace{2cm}\times
       \sqrt{ \big[ m_\psi^2 - (m_\chi + m_S(T))^2 \big]
              \big[ m_\psi^2 - (m_\chi - m_S(T))^2 \big] } \,,
  \label{eq:kin:psi-chi-S} \\[0.2cm]
  \Gamma(S \to \chi \bar\psi)
    &= \frac{y_{\psi\chi}^2}{8 \pi}
       \frac{m_S^2(T) - (m_\chi + m_\psi)^2}{m_S^3(T)}
				\nonumber\\
    &\hspace{2cm}\times
       \sqrt{ \big[ m_S^2(T) - (m_\chi + m_\psi)^2 \big]
              \big[ m_S^2(T) - (m_\chi - m_\psi)^2 \big] } \,,
  \label{eq:kin:S-chi-chi} \\[0.2cm]
  \sigma(S \psi \to S \chi)
    &\simeq \frac{y_\psi^2 y_{\psi\chi}^2}{32\pi s^2 (s - m_\chi^2)^3}
       \Big[ (s - m_\chi^2) (5 s^3 + 55 m_\chi^2 s^2 + 3 m_\chi^4 s + m_\chi^6)
                                \nonumber\\
    &\hspace{5cm}
           - 2 s^2 ( s^2 - 18 m_\chi^2 s - 15 m_\chi^4) \log\Big( \frac{m_\chi^2}{s} \Big)
       \Big] \,,
  \label{eq:kin:sigma-S-psi-S-chi} \\[0.2cm]
  \sigma(S S \to \chi \bar\psi) \big|_{v_S = 0}
    &\simeq \frac{y_\psi^2 y_{\psi\chi}^2}{8 \pi s^3}
      \bigg[ 2 (s^2 + 16 m_\chi^2 s - 32 m_\chi^4)
        \log\bigg( \frac{s + \sqrt{s (s - 4 m_\chi^2)}}
                        {s - \sqrt{s (s - 4 m_\chi^2)}} \bigg)
                                \nonumber\\
    &\hspace{5cm}
      - 4 (s + 8 m_\chi^2) \sqrt{s (s - 4 m_\chi^2)} \bigg] \,.
  \label{eq:kin:sigma-S-S-chi-psi}
\end{align}
In the last two expressions, we have taken the limit $m_\psi \simeq m_\chi$
and $m_S \simeq 0$. Moreover, we have set the width of $\psi$ to $\Gamma_\psi = 0$.
In \cref{eq:kin:sigma-S-S-chi-psi}, we have also set $v_S = 0$ because the
full expression is fairly lengthy, and we will see below that the channel
$S S \to \chi \bar\psi$ is very subdominant when $v_S \neq 0$.  In our numerical
analysis below, we of course use the full expressions.


It is important that, thanks to the non-zero $\ev{S}$ at low temperatures, $S$
can decay through its Higgs portal coupling $\lambda_{p4}$ (or also through
$\lambda_{p3}$).  If this decay was not present, $S$ would have a relic abundance
that would be too large. For $m_S$ below the $W$ and $Z$ thresholds, the decay rate is
\begin{align}
  \Gamma(S \to f \bar{f}) &=
    \sum_f \frac{y_f^2 \, \theta_{hS}^2(T) \, m_S(T)}{8\pi}
             \bigg( 1 - \frac{4 m_f^2}{m_S^2(T)} \bigg)^{3/2}
             \theta\big( m_S(T) - 2 m_f \big)\,.
  \label{eq:kin:Gamma-S-f-f}
\end{align}
Here, the sum runs over light fermions, $f = e,\,\mu,\,\tau,\,u,\,d,\,s,\,c,\,b$,
and $y_f = \sqrt{2} m_f / v_H$ are the corresponding Yukawa couplings.
The $h$--$S$ mixing angle is
\begin{align}
  \tan\theta_{hS}(T) = \frac{(\lambda_{p3} \mu_S + \lambda_{p4} v_S(T)) v_H(T)}
                            {m_H^2(T) - m_S^2(T)} \,.
  \label{eq:kin:theta}
\end{align}
In these expressions $v_S(T)$ and $v_H(T)$ are the $S$ and Higgs vevs, respectively.
$S$ also couples to SM particles via annihilations. After the
electroweak phase transition, the main annihilation process is $S S \to h^* \to
f \bar{f}$, with cross section
\begin{align}
  \sigma v_\text{rel}(S S \to f \bar{f})
    &= \sum_f \frac{C_f y_f^2 \lambda_{p4}^2 v_H^2(T)}{32 \pi (m_H^2(T) - 4 m_S^2(T))^2}
       \bigg( 1 - \frac{m_f^2}{m_S^2(T)} \bigg)^{3/2}
       \theta\big( m_S(T) - m_f \big) \,
  \label{eq:kin:sigma-v-S-S-f-f}
\end{align}
in the non-relativistic limit.  In this expression, $C_f$ is a colour factor.
We will choose $\lambda_{p4} \simeq 10^{-3}$, which makes $S$ decays, inverse decays, and
annihilations fast enough to keep $S$ in thermal equilibrium with the SM at
all $T \lesssim m_\chi$, where DM freeze-in dominantly occurs. We have verified
that this is always possible for $m_\chi > m_S \gtrsim 3$\,GeV. If $S$ were not
in thermal equilibrium during DM freeze-in, the model would not be invalidated,
but the dark and visible sector temperatures would differ, complicating the
analysis.

We will also assume that, while DM freezes in, $\psi$ remains in thermal
equilibrium with the SM sector, either through $\psi \bar\psi \leftrightarrow S
S$ or through interactions with a second new scalar $S'$ as explained below
\cref{eq:kin:V}.  The cross section for $\psi \bar\psi \to S S$ in the
non-relativistic limit is
\begin{align}
  \sigma v_\text{rel}(\psi \bar\psi \to S S)
    &= \frac{v_\text{rel}^2 y_\psi^4}{24 \pi}
       \frac{\sqrt{m_\psi^2 - m_S^2(T)} \, \big(2 m_S^4(T) m_\psi - 8 m_S^2 m_\psi^3
                  + 9 m_\psi^5 \big)}{(m_S^2(T) - 2 m_\psi^2)^4} \,,
  \label{eq:kin:sigma-v-psi-psi-S-S}
\end{align}
where $v_\text{rel}$ is the relative velocity of the annihilating $\psi$
particles. Note the $v_\text{rel}^2$ suppression of the annihilation cross
section. Eventually, $\psi$ will freeze out, and we ensure that this
happens late enough for its relic abundance to make up a subdominant contribution
to the DM density (less than $1\%$).  Note that $\psi$ is not absolutely stable, but can
decay even at $T=0$ through $\psi \to \chi (S^*/H^* \to f \bar{f})$. The rate
of this decay is (for $m_\psi - m_\chi > m_S$, $(m_\psi - m_\chi)/m_\chi \ll 1$,
and $m_f = 0$) given by
\begin{align}
  \Gamma(\psi \to \chi f \bar{f}) &\simeq
    \sum_f \frac{C_f y_f^2 y_{\psi\chi}^2 \theta_{hS}^2(T)}{120 \pi^3}
             \frac{m_\psi^5(m_h^2 + m_S^2)^2}{m_h^4 m_S^4}
             \bigg( 1 - \frac{m_\chi}{m_\psi} \bigg)^5 \,.
  \label{eq:Gamma-psi-f-f}
\end{align}
Here, the sum runs over all kinematically accessible SM fermion species.  Since
we treat fermions as massless, \cref{eq:Gamma-psi-f-f} will not be accurate
near any of the fermion mass thresholds.  Even though $\psi$ freezes out with
only a subdominant relic abundance, its decays may violate constraints if they
happen around the time of recombination \cite{Cadamuro:2010cz, Cadamuro:2011fd,
Cadamuro:2012rm}. We therefore demand $\tau_\psi \equiv 1/\Gamma_\psi \gtrsim
10^{11}$\,sec or $\tau_\psi \lesssim 10^5$\,sec~\cite{Cadamuro:2011fd}.  We
have verified that for the parameter region we will discuss in the following,
$\lambda_{p4}$ and thus $\theta_{hS}(T)$ can indeed be adjusted such that
$\tau_\psi \gtrsim 10^{11}$\,sec.  Even in this case, care must be taken that
the residual abundance of $\psi$ at the time of decay is tiny ($\lesssim
10^{-10} \times (\tau_\psi / 10^{15}\,\text{sec})$ times the DM abundance) to
avoid anomalous reionization \cite{Slatyer:2016qyl}. We have checked that this
can be automatically achieved for DM masses $\lesssim 1$~GeV. In this
case, $\tau_\psi$ is large because the only decay channels available to $\psi$
are suppressed by the small Yukawa couplings of light quarks and leptons.  At
larger DM mass, the easiest way of ensuring compatibility of the model with
reionization constraints is to introduce an auxiliary fermion $\chi'$, with
$m_\chi' \ll m_\chi$, and with a Yukawa coupling to $S$ and $\psi$ chosen such
that the decay $\psi \to \chi' f \bar{f}$ is much slower than freeze-in of
$\chi$, but still occurs significantly before recombination ($\tau_\psi
\lesssim 10^5$\,sec~\cite{Cadamuro:2011fd}).

\subsection{The Effective Potential}
\label{sec:kin:Veff}

To correctly describe the evolution of the scalar sector of the model from
\cref{eq:kin:L,eq:kin:V} in the hot early Universe, we must go beyond the
tree level potential and consider the finite temperature effective potential $V^\text{eff}$
which includes radiative corrections and thermal effects.
Since we assume the
portal couplings $\lambda_{p3}$ and $\lambda_{p4}$ to be small, we can treat
the evolution of the dark sector potential as independent from that of the
visible sector potential
(we will consider the case of large portal couplings in \cref{sec:vev,sec:mix}).
We begin with the approximate tree level potential in the dark sector,
\begin{align}
  V^\text{tree}(S) &\approx -\frac{\mu_S^2}{2} S^2 + \frac{\lambda_{S4}}{4!} S^4 \,.
\end{align}
The effective potential $V^\text{eff}$ is defined in the usual way
\cite{Peskin:1995ev}: one first rewrites the generating functional, $E[J] = i
\log Z[J]$,  as a functional of $v_S(x)$ instead of the external source field
$J(x)$. (Here, $Z[J]$ is the partition function.) This is achieved by relating
$J(x)$ with $v_S(x)$ via $v_S(x) = \delta E[J] / \delta J$. Note that, in the
presence of an $x$-dependent external source, $v_S(x)$ becomes a function of
$x$ as well.  The effective action $\Gamma^\text{eff}$, which is in turn the
spacetime integral of the effective potential $V^\text{eff}$, is given by a
Legendre transform: $\Gamma^\text{eff}[v_S] = \int d^4x \, V^\text{eff}[v_S] \equiv -E[J]
- \int d^4y \, J(y) \, v_S(y)$. We see that $\Gamma^\text{eff}$ has the
property that $\delta \Gamma^\text{eff}[v_S] / \delta v_S = 0$, that is, the
vacuum configuration $v_S$, including all quantum corrections, is obtained from
$\Gamma^\text{eff}$ using a variational principle. Of course, $\Gamma^\text{eff}$
itself needs to be computed perturbatively.

As we outline in more detail in \cref{sec:app-eff-pot}, the leading corrections
that distinguish $V^\text{eff}$ from $V^\text{tree}$ are the Coleman-Weinberg
term $V^\text{CW}$ that corresponds to resummed 1-loop diagrams at
$T=0$~\cite{Coleman:1973jx}, the thermal one-loop
contribution~\cite{Dolan:1973qd}, $V^T$, and the resummed series of higher
order ``daisy'' diagrams, $V^\text{daisy}$~\cite{Carrington:1991hz,
Quiros:1999jp, Ahriche:2007jp, Delaunay:2007wb}. With our assumption that
$y_{\psi\chi}$, $\lambda_{p3}$, and $\lambda_{p4}$ are tiny, loops involving
$H$ and $\chi$ are negligible. Loops involving $\psi$ could be relevant at
temperatures not too far below $m_\psi$ if $y_\psi \gtrsim 0.01$. In the
following we will assume $y_\psi \lesssim 0.01$ to simplify the analysis.
As explained
below \cref{eq:kin:V} this will require a different mechanism for keeping
$\psi$ in thermal equilibrium throughout DM freeze-in. We have
verified that our toy model can be phenomenologically successful also for
larger $y_\psi$.
For $y_\psi \lesssim 0.01$, the only relevant diagrams contributing to
$V^\text{eff}$ are those involving the quartic coupling $\lambda_{S4}$. In
other words, the sums in \cref{eq:app:VCW,eq:app:VT,eq:app:Vdaisy} run only
over $S$.  The coefficient $n_i$, which can be interpreted as counting degrees
of  freedom (although see~\cite{Delaunay:2007wb}), is $n_S = 1$.
As a function of the field value, the mass of $S$ is given by
\begin{align}
  m_S^2  &\simeq  -\mu_S^2  +  \frac{\lambda_{S4}}{2} S^2\,.
  \label{eq:kin:mS}
\end{align}
$V^\text{daisy}$ also depends on the thermal, or Debye, mass,
which is given by the 1-loop 
self energy
at non-zero $T$. The Debye mass of $S$ is given by~\cite{Carrington:1991hz}
\begin{align}
  \Pi_S(T) &= \frac{T^2}{24} (\lambda_{S4} + 4 y_\psi^2) \,.
  \label{eq:kin:thermalmass}
\end{align}


\begin{figure}
  \begin{center}
    \begin{tabular}{cc}
      \includegraphics[width=0.45\textwidth]{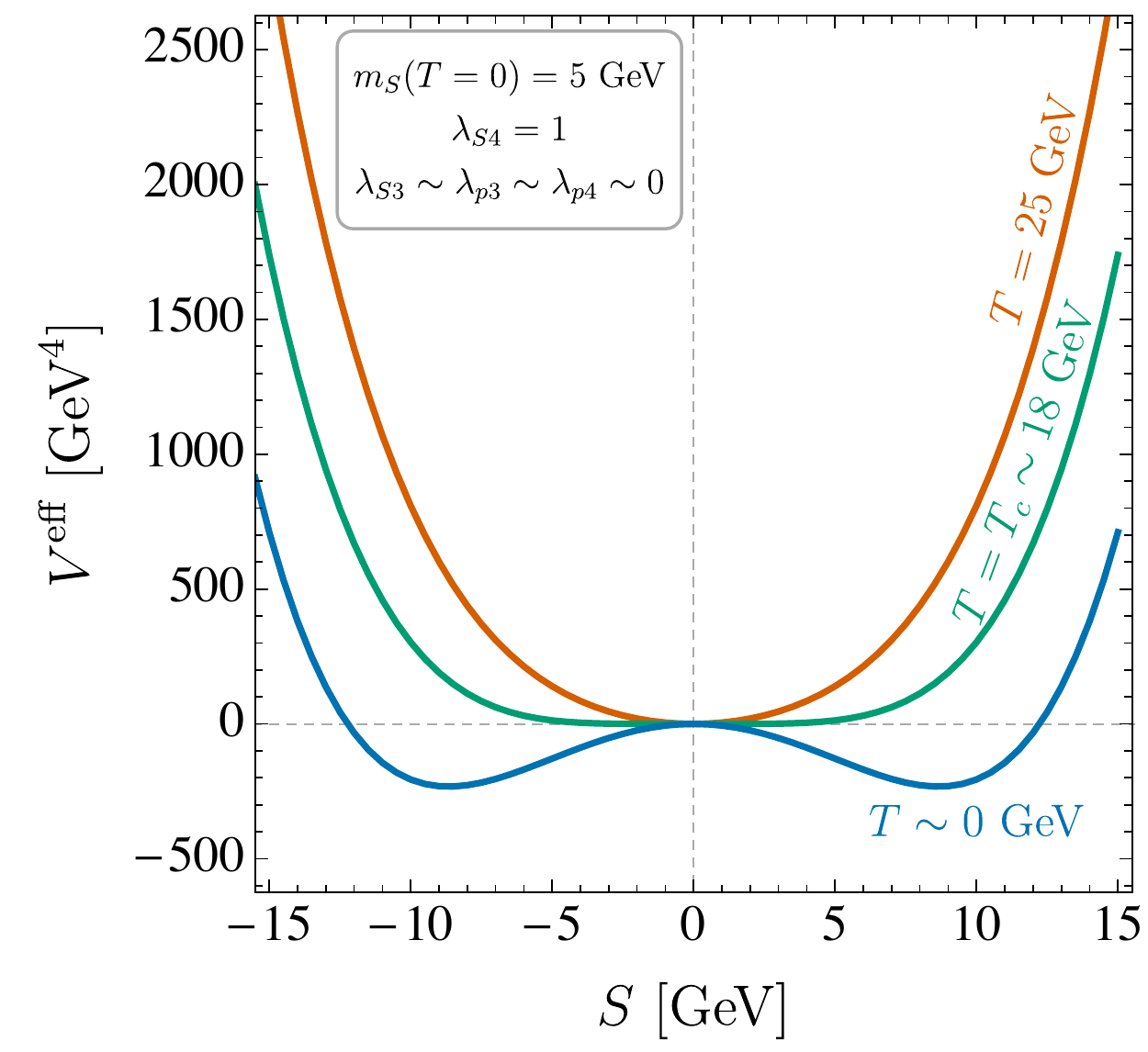} & 
      \includegraphics[width=0.45\textwidth]{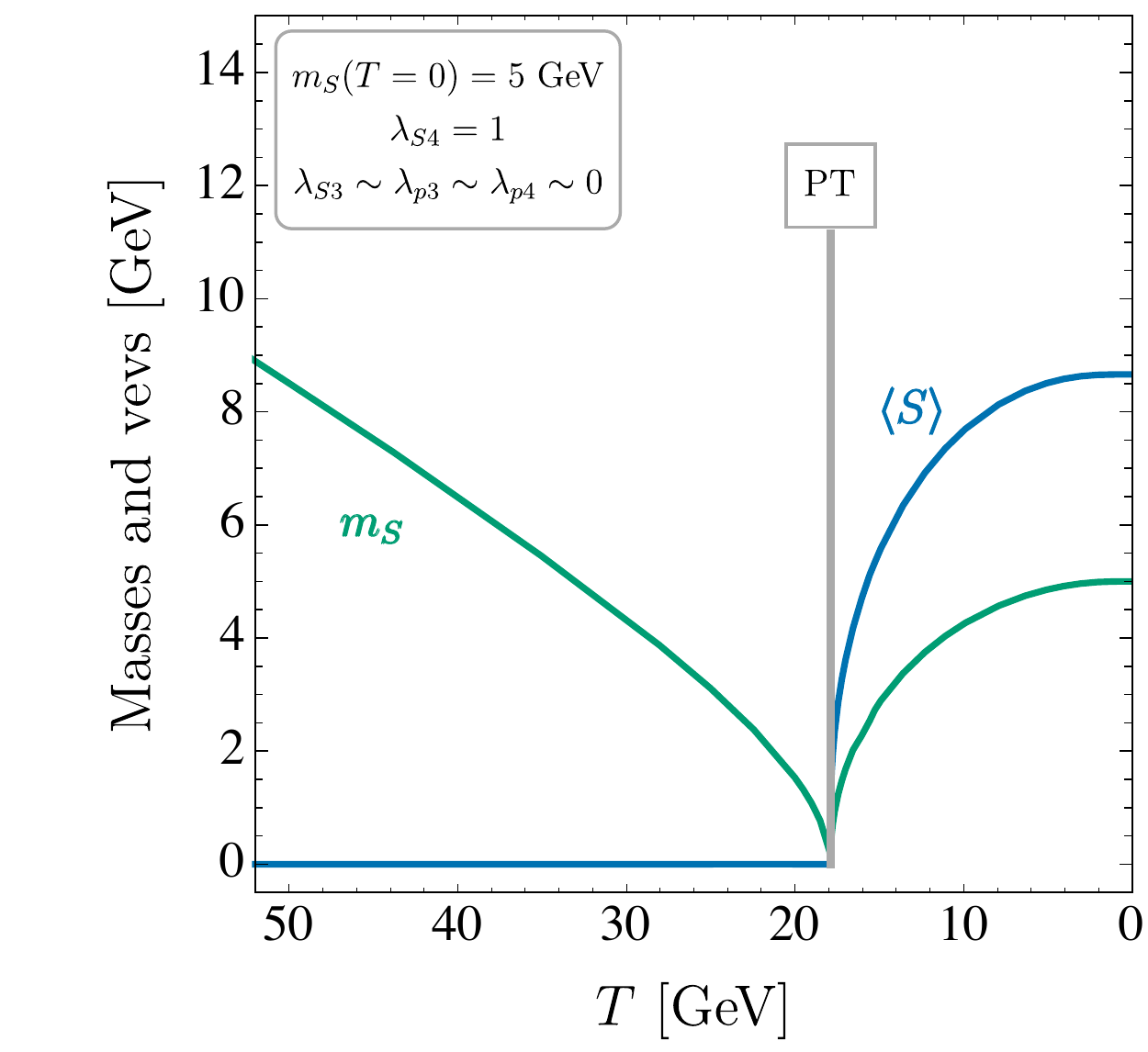}
    \end{tabular}
  \end{center}
  \caption{The new scalar effective potential at a relatively high temperature, at $T=T_c$ 
  and at $T\sim0\,\text{GeV}$, (left).  A constant term has been subtracted at each $T$ so that
   $V^\text{eff}(0)=0$.  The evolution of the new scalar mass and vev with temperature, (right), 
   with the temperature of the phase transition indicated.
   Note that we neglect contributions from $\psi$ loops to $V^\text{eff}$,
   assuming $y_\psi \lesssim 0.01$.
   }
  \label{fig:sec2-eff-pot}
\end{figure}


With the effective potential in hand, we can now consider the evolution of
$m_S$ and $v_S \equiv \ev{S}$ as a function of $T$.  This allows us to describe the
phase transition, which is analogous to the electroweak phase transition in the
SM.  We use the program \texttt{CosmoTransitions}~\cite{Wainwright:2011kj,
Kozaczuk:2014kva, Blinov:2015sna, Kozaczuk:2015owa} to track the minimum of the
effective potential, to find $\ev{S}$ as a function of $T$, and to determine
the mass of $S$ as the second derivative of $V^\text{eff}(S)$.  Although in our
particular toy model it would be easy to do the computation without invoking
\texttt{CosmoTransitions}, we still use it for consistency with
\cref{sec:vev,sec:mix}.

The effective potential at several temperatures is shown in
\cref{fig:sec2-eff-pot} (left), while the behaviour of $\vev{S}$ and $m_S(T)$ is
shown in \cref{fig:sec2-eff-pot} (right).  In the left panel we see the well known
behaviour of a second order phase transition or cross-over: at high
temperatures, $T \gg T_c$, the effective potential has its
minimum at $S=0$.  At the critical temperature, $T_c$, the minimum begins to
move away from $S=0$ and a non-zero vev begins to develop.  
At the present time, near $T=0\, \text{GeV}$, the
effective potential has its minimum at $S\simeq 9\,\text{GeV}$.
In \cref{fig:sec2-eff-pot} (right), we similarly see that at high temperatures, $S$
has no vev, and its effective mass is large thanks to thermal corrections $\propto
T$.  As the temperature drops, the mass of $S$ drops as well and approaches
zero at the phase transition.  This behaviour can be understood also from the
green curve in \cref{fig:sec2-eff-pot} (left): at the transition temperature, its
second derivative at the minimum is zero.  After the phase transition, the vev
and the mass of $S$ grow and quickly approach their present day values, 
$m_S(T=0) = 5\,\text{GeV}$ and $\ev{S} \simeq 9\,\text{GeV}$.

As well as this temperature dependence of the $S$ mass, the fermions 
may also have temperature dependent mass contributions.  Since we 
take the tree level fermion masses to be much larger than the $S$ mass, 
self-energy diagrams evaluated at $T \neq 0$ will only give a small contribution 
near the phase transition
(fermions have no zero Matsubara mode, so the self-energy contributions 
for fermions are smaller than those for bosons).
However, if $y_\psi$ is large, then there can be significant corrections to the 
$\psi$ mass from the Lagrangian term $y_\psi \bar{\psi}S \psi$ when $S$ obtains 
its vev.  The $T$ dependent $\psi$ mass is then
\begin{align}
m_\psi (T) =&\, m_\psi - y_\psi \ev{S}.
\end{align}

The crucial point for us is that, thanks to the behaviour of $m_S(T)$ and $m_\psi(T)$,
the freeze-in channel $\psi \to \chi S$ is kinematically closed long before
and long after the dark sector phase transition, while around the
transition temperature, it is open and DM freeze-in can proceed efficiently.

\subsection{Results and Discussion}
\label{sec:kin:results}

\begin{figure}
  \begin{center}
    \begin{tabular}{cc}
      \includegraphics[width=0.45\textwidth]{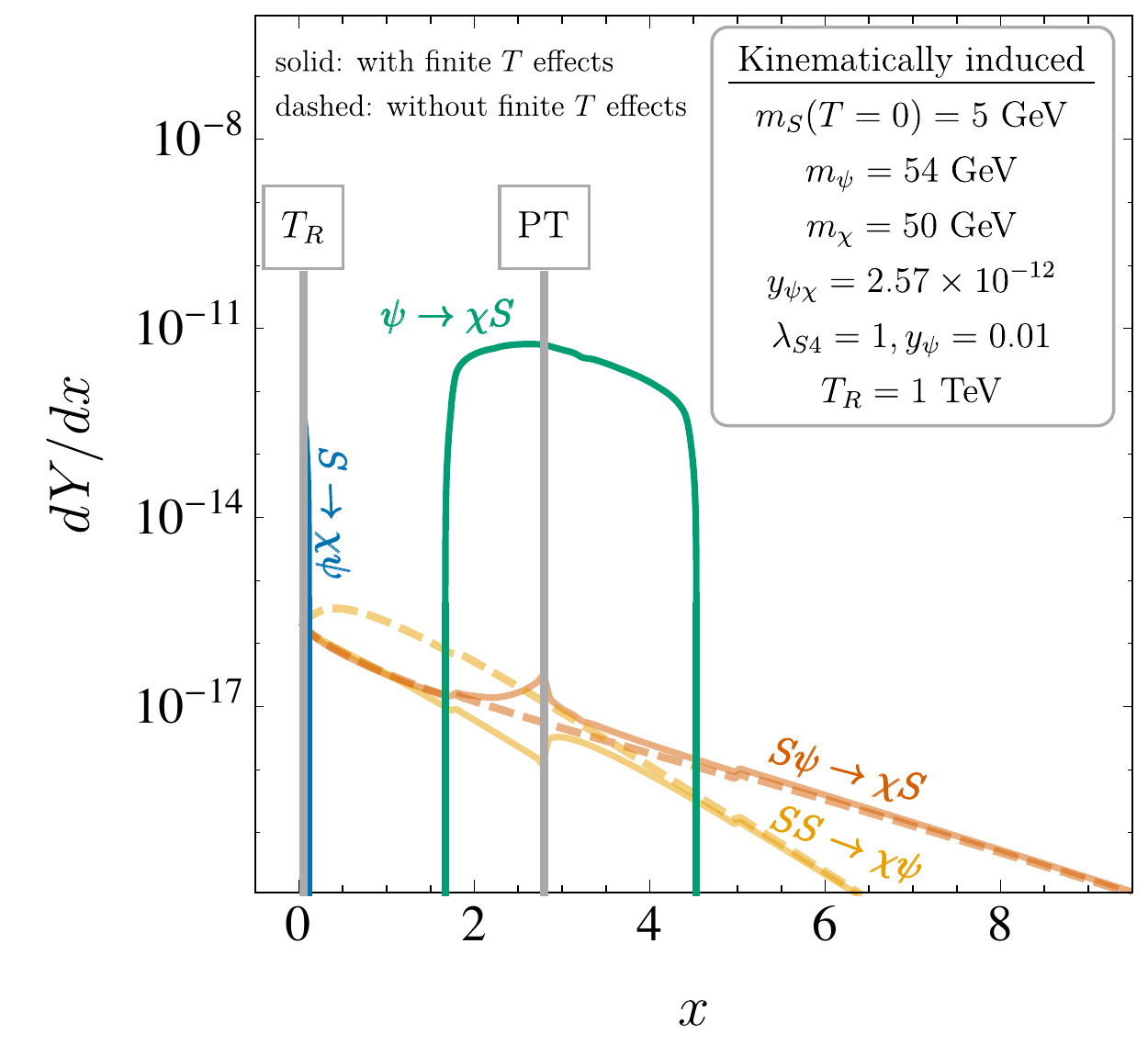} &
      \includegraphics[width=0.45\textwidth]{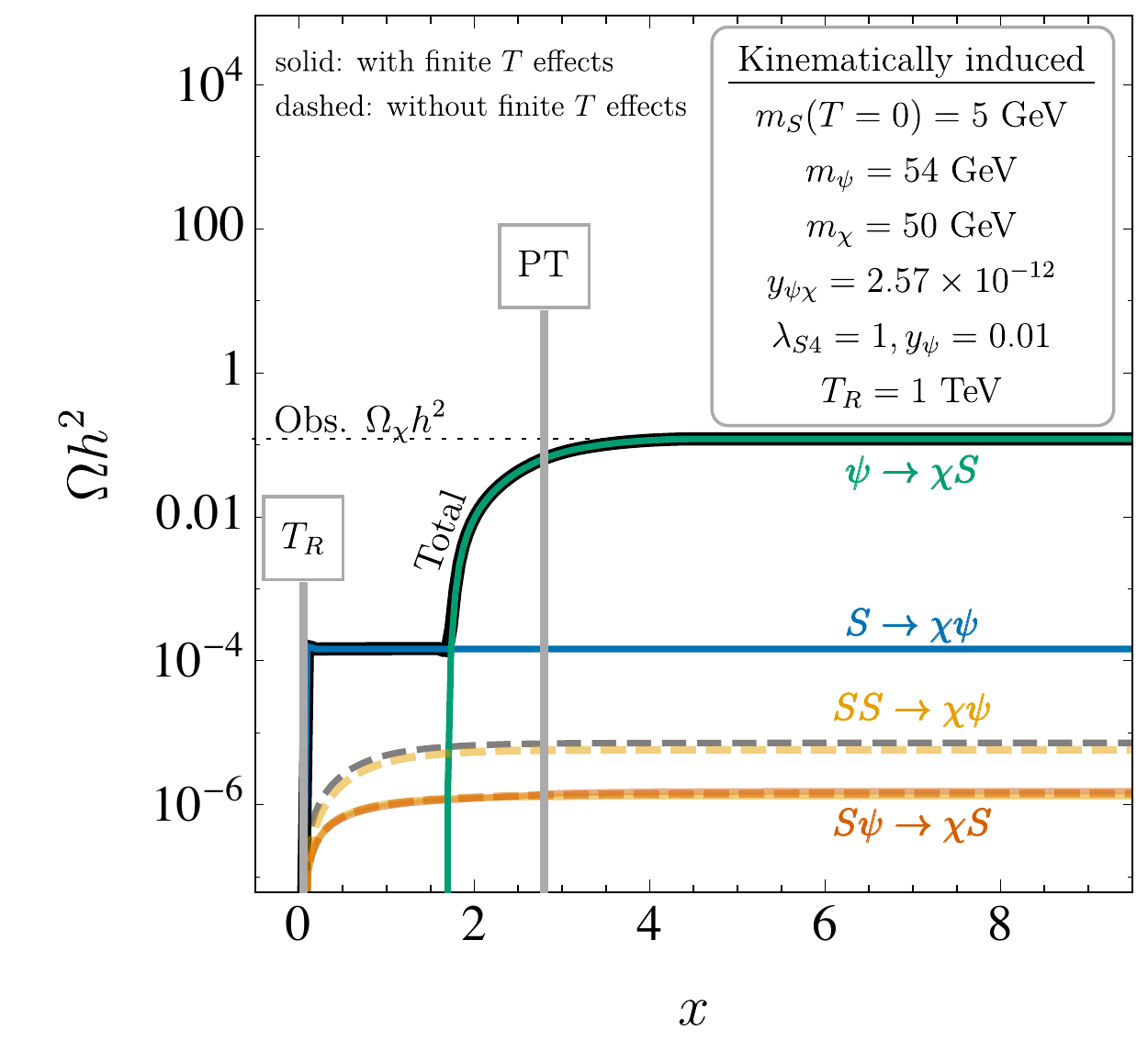} \\[-0.2cm]
      \includegraphics[width=0.45\textwidth]{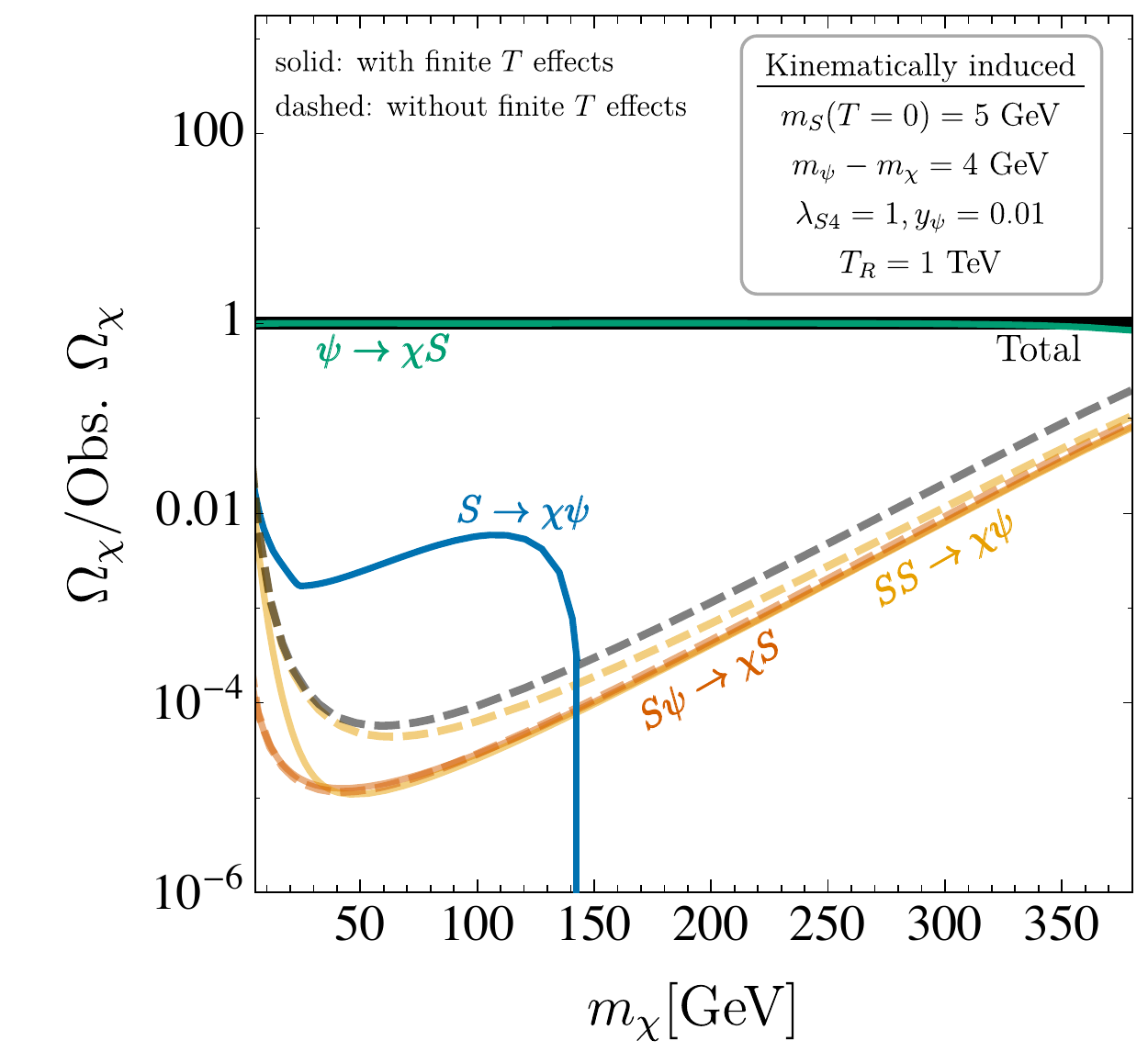} &
      \includegraphics[width=0.45\textwidth]{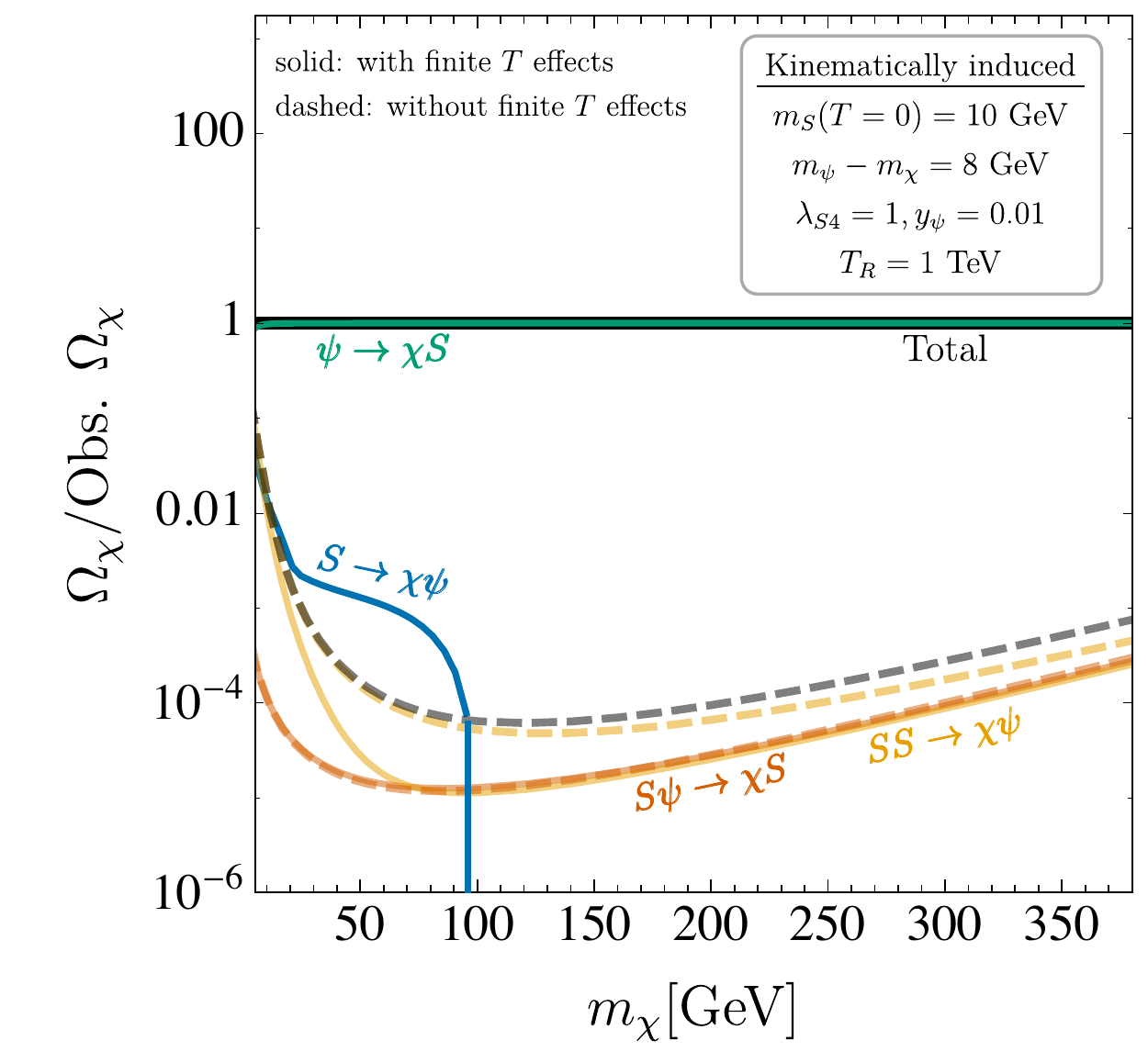} \\[-0.2cm]
    \end{tabular}
  \end{center}
  \caption{
    The instantaneous change in the yield (top-left) and the resulting freeze-in 
    relic abundance extrapolated to the present day (top-right) for our benchmark point, including (solid lines) 
    and ignoring (dashed lines) finite temperature effects.  Also shown 
    is the relative final abundance (normalised to the abundance when finite temperature 
effects are included) through each channel as a function of $m_\chi$ 
    for $m_S(T=0) = 5\,\text{GeV}$ (bottom-left) and $m_S(T=0) = 10\,\text{GeV}$ (bottom-right).
  }
  \label{fig:vev-mh-T}
\end{figure}

In \cref{fig:vev-mh-T} (top-left) we show the instantaneous DM yield from each
freeze-in channel in \cref{fig:kin:diagrams}, including (solid lines) and
ignoring (dashed lines) finite temperature contributions to the scalar masses
and vevs and to the fermion masses.  If we ignore the finite temperature corrections, only two channels
($S\psi \to \chi S$ and $SS \to \chi \psi$) contribute to the $\chi$ abundance.
These contributions are largest at high temperatures (or small $x$, where $x =
m_\chi / T$), where the abundance of $\psi$ is not Boltzmann suppressed and
where $S$ have enough energy to produce the heavier states $\psi$ and $\chi$.
The instantaneous freeze-in yield reduces smoothly as the temperature reduces,
except for small steps where SM particles freeze-out and the effective number
of relativistic degrees of freedom in the Universe, $g_\text{eff}$, changes.


If finite temperature effects are included, then two new channels contribute at
different times.  At very high temperatures, $S$ has a sufficiently large mass
that it can decay to $\chi \bar{\psi}$.  As $\chi$ and $\psi$ are much heavier than
$S$ at $T=0\,\text{GeV}$, this channel is only open at very high temperatures
($ x \lesssim 0.1 $), and it no longer contributes at lower
temperatures.  As the Universe approaches the phase transition in the dark
sector, the mass of $S$ reduces until it becomes smaller than $m_\psi(T) - m_\chi$
at $x \simeq 1.7 $ (the mass of $\psi$ is constant before $S$ obtains its vev).  
At this point, the decay $\psi \to \chi S$
becomes kinematically possible and $\chi$ is produced.  This happens at a rate
much larger than that via the  $S\psi \to \chi S$ and $S S \to \chi \psi$
channels because the latter channels are suppressed by the off-shellness of the
intermediate $\psi$ propagator.  The $\psi \to \chi S$ channel reaches its
maximum rate around the dark phase-transition, where $m_S(T)$ goes to zero.  As
the temperature further reduces, the mass of $S$ increases and the 
mass of $\psi$ reduces until the channel closes at $x \simeq 4.5$.
Comparing the rates with and without including finite temperature effects,
we see that these effects are relevant in all channels.  The rate of $S \psi \to
\chi S$ shows a peak at the dark phase-transition because the
intermediate $s$-channel $\psi$ propagator in the third diagram of
\cref{fig:kin:diagrams} can go nearly on-shell when $m_S(T)$ is small around
the transition temperature.  This is essentially a manifestation of the
infrared divergence of the corresponding amplitude.

The resulting relic abundance of $\chi$ 
extrapolated to zero redshift is shown in \cref{fig:vev-mh-T} (top-right). The extrapolated
abundance at a given $x \equiv m_\chi / T$ is obtained by rescaling the number density
at this time by the subsequent expansion of the Universe and normalising
to the critical density today.
Here we clearly see that the dominant contribution to the relic abundance comes 
from the $\psi \rightarrow \chi S$ channel.  We emphasise that if finite temperature effects were not 
included, the calculated relic abundance would be incorrect by a factor $\mathcal{O}(10^4)$.
For the benchmark parameters chosen, the resulting
$\chi$ abundance matches the observed relic abundance for $y_{\psi\chi}
= 2.57 \times 10^{-12}$.  

In \cref{fig:vev-mh-T} (bottom-left) we show the abundance produced through each channel as 
a function of $m_\chi$, normalised to the abundance when finite temperature 
effects are included.  We keep the mass difference between $\psi$ at $T=0$ and $\chi$ fixed 
at $4\,\text{GeV}$.  The fraction of $\chi$ produced through channels other than
$\psi \to \chi S$ is always small, but it is smallest
for $m_\chi \simeq 50\,\text{GeV}$.  
At lower values of $m_\chi$, $S$ no longer requires significant thermal energy to produce $\chi$ 
via $S\psi \to \chi S$ and $SS \to \chi \psi$, so the $\psi \to \chi S$ 
channel produces a smaller fraction of $\chi$.  As the value of $m_\chi$ increases, 
processes which occur at lower temperatures receive more Boltzmann suppression than 
those occurring at higher temperatures.
This means that the amount 
of $\chi$ produced through $S\psi \to \chi S$ and $SS \to \chi \psi$ (which is important at 
high temperatures) is mildly reduced whereas the amount produced via $\psi \to \chi S$ 
(which is important around the phase transition) has greater Boltzmann suppression, reducing its relative 
importance.

Finally, \cref{fig:vev-mh-T} (bottom-right) shows the freeze-in abundance of $\chi$ 
through the different channels for $m_S = 10\,\text{GeV}$, where now the phase 
transition occurs at $T = 36\,\text{GeV}$.  We see that the picture is qualitatively
similar to \cref{fig:vev-mh-T} (bottom-left). For $m_s = 10\,\text{GeV}$,
there is a milder reduction in the 
relative importance of the $\psi \to \chi S$ channel at large $m_\chi$, 
due to a milder Boltzmann suppression of the $\psi$ abundance at the 
phase transition.  In both \cref{fig:vev-mh-T} (bottom-left) and (bottom-right), 
the yields due to $S\psi \to \chi S$
are similar including or ignoring finite temperature corrections.  This 
channel predominantly produces $\chi$ at high temperatures, where the 
particle momentum dominates over the particle masses.  
For the small Yukawa coupling $y_\psi = 0.01$ and large quartic coupling $\lambda_{S4} = 1$, 
the $SS \to \chi \psi$ production 
rate is much larger when $S$ has a vev, which explains the difference seen between 
the curves that include or ignore the finite temperature corrections in this channel.

For definiteness in our numerical calculations, we fix a reheating 
temperature $T_R = 1\,\text{TeV}$ which is sufficiently large so 
that any freeze-in at higher temperatures will produce negligible 
abundance of $\chi$.

We finish this section by noting a particularly simple model which shares 
many features with the toy model discussed above.  If the SM is extended 
by two dark sector fermions, a SM gauge singlet, $\chi$, and an $su(2)_L$ doublet 
with hypercharge, $\psi$, then the SM Higgs can play the role of $S$ above.  The 
Lagrangian term $\bar{\psi} H \chi + h.c.$ can lead to processes which produce $\chi$ via freeze-in.  
If $m_\psi - m_\chi < m_h(T=0)$ and $m_\chi < m_\psi$, then the channel 
$\psi \to \chi H$ will be open only when the mass of the SM Higgs is reduced 
during the second order electroweak phase transition.  For 
$\psi$ lighter than 1.1\,TeV, it freezes-out as a subdominant 
component of the dark matter abundance~\cite{Cirelli:2005uq}.  The remaining relic abundance can 
then be provided by the freeze-in of $\chi$.  The calculation of the abundance 
is somewhat complicated and we defer this to later work.

\section{Vev-Induced Production with a Vev Flip-Flop}
\label{sec:vev}

In the previous section we have highlighted the potential importance of
including finite temperature corrections to particle masses in calculations of DM
freeze-in.  In this section, we consider a model with a fermionic DM candidate
$\chi$ and with a scalar sector identical
to the one in \cref{eq:kin:L}, but focussing on a different region of
parameter space. Namely, we now take the Higgs portal couplings so large that
a two-step phase transition (or ``vev flip-flop''~\cite{Baker:2016xzo}) is
realised \cite{Profumo:2007wc, Cline:2009sn, Espinosa:2011ax, 
Cui:2011qe, Cline:2012hg, Fairbairn:2013uta, Curtin:2014jma, Beniwal:2017eik}
(see also \cite{Cohen:2008nb}).
  In other words, the Universe goes through a phase where the new
scalar $S$ obtains a non-zero vev, but this vev jumps back to zero in a
first order electroweak phase transition.  The value of $\ev{S}$ will control
the DM freeze-in rate, so DM can only be efficiently produced during a relatively
short time interval. The final DM abundance is determined not only by the
relevant coupling constants, but also by the length of this time interval.
We dub this mechanism ``vev-induced production.''

\subsection{Toy Model}
\label{sec:vev:model}

The field content of the dark sector in this toy model is shown in
\cref{tab:vev:particles}.  As in \cref{sec:kin}, our dark matter candidate is
a Dirac fermion, $\chi$, which is a SM gauge singlet.  We assume that it is
stabilised by a $\mathbb{Z}_2$ symmetry.
The DM mass $m_\chi$ and the scalar mass parameter $|\mu_S|$
are taken to be $\simeq 100$\,GeV.  The relevant terms in the Lagrangian are
\begin{align} 
  \mathcal{L} &\supset  y_\chi S \bar{\chi} \chi  -  V(H,S) \,,
  \label{eq:vev:L}
\end{align}
with $V(H, S)$ again given by \cref{eq:kin:V}.  DM will freeze-in
via the Yukawa coupling $y_\chi$; consequently, this coupling needs to be
tiny.  As in \cref{sec:vev:model}, this could be motivated in extra-dimensional
scenarios by localising $\chi$ far away from the other fields along a
fifth dimension.
$\lambda_{S3}$ and $\lambda_{p3}$ are assumed to be small as well to simplify the
analysis. Note that an
extra global $\mathbb{Z}_2$ symmetry is restored if $y_\chi$, $\lambda_{S3}$, and
$\lambda_{p3}$ are set to zero. Therefore, small values for these couplings
are natural in the 't~Hooft sense~\cite{tHooft:1979rat}.
Setting $\lambda_{p3} \ll 1$ means we can ignore mixing between the SM Higgs
boson and $S$ at $T=0$.  In this limit, the mass of $S$ is given by
\begin{align}
  m_S^2(T=0) = -\mu_S^2 + \frac{\lambda_{p4}}{2} v_H^2(T=0)
  \label{eq:vev:mS}
\end{align}
at tree level and $T=0$. We see that for $\mu_S \simeq v_H(T=0)$ and
$\lambda_{p4} \simeq 1$, a situation can be realised where $\lambda_{p4} v_H^2(T=0) / 2
> \mu_S^2$. In this case, $v_S(T) \equiv \ev{S} \ne 0$
as long as $v_H(T) = 0$ (barring thermal corrections for the moment), but when
$v_H(T)$ becomes significantly different from zero, the term quadratic in
$S$ in the scalar potential
experiences a sign flip, making $v_S(T) = 0$ energetically favourable in the broken
phase of electroweak symmetry.  $v_S(T) = 0$ is also realised at very
early time thanks to thermal corrections to $V^\text{eff}$. These corrections
are large especially when $\lambda_{S4} \gtrsim 1$.  This behaviour is the gist
of the vev flip-flop, and it defines the parameter region we will be interested
in in the following: $\mu_S \simeq v_H(T=0)$, $\lambda_{p4} \simeq\lambda_{S4} \simeq 1$.


\begin{table}
  \centering
  \begin{minipage}{10cm}
    \begin{ruledtabular}
    \begin{tabular}{ccccc}
      Field  & Spin          & $\mathbb{Z}_2$ & mass scale \\
      \hline
      $S$    & $0$           & $+1$           & $m_S(T=0) \simeq 100~\text{GeV}$ \\
      $\chi$ & $\frac{1}{2}$ & $-1$           & $m_\chi \simeq100~\text{GeV}$ \\
      \bottomrule
    \end{tabular} 
    \end{ruledtabular}
  \end{minipage}
  \caption{The dark sector of the vev-induced freeze-in scenario.}
  \label{tab:vev:particles}
\end{table}


\subsection{The Effective Potential \& The Vev Flip-Flop}
\label{sec:vev:Veff}

To quantitatively compute the effective potential for the model defined in
\cref{eq:vev:L}, the same methods as in \cref{sec:kin:Veff} can be applied, but
since the Higgs portal coupling $\lambda_{p4}$ is no longer negligible, we
need to consider the joint evolution of the visible and dark scalar sector. In
other words, we need to treat $V^\text{eff}$ as a function of both $S$ and $H$.
As explained in \cref{sec:kin:Veff,sec:app-eff-pot}, the one-loop contribution
to the effective potential, $V^\text{CW}(h,S) + V^T(h,S)$, depends on the field
dependent masses of all particles with couplings to the scalars.  In the model
from \cref{eq:vev:L}, the field dependent masses of $W^i, B$ and $t$ are the
same as in the SM. In particular the gauge boson mass eigenvalues are
$m_{W^\pm}^2 = \tfrac{1}{4} g^2 h^2$, $m_Z^2 = \tfrac{1}{4} (g^2 + g'^2) h^2$,
$m_\gamma^2 = 0$, and $m_t = y_t h / \sqrt{2}$. Here, $g$ and $g'$ are the
$su(2)_L$ and $u(1)_Y$ gauge couplings, respectively, and $y_t$ is the top
quark Yukawa coupling.  The mass matrix of the neutral CP-even Higgs bosons is
\begin{align}
  \hspace{-0.3cm}
  m_{(h,S)}^2(h,S) &=
  \begin{pmatrix}
    -\mu_H^2 + \lambda_{p3}\,\mu_SS + \frac{1}{2}\lambda_{p4}\,S^2 + 3\lambda_{H4} \,h^2 &
      (\lambda_{p3}\,\mu_S + \lambda_{p4}\,S) h \\
    (\lambda_{p3}\,\mu_S + \lambda_{p4}\,S) h &
    -\mu_S^2 + \lambda_{S3}\,\mu_S\,S + \frac{1}{2}\lambda_{S4}\,S^2
                                      + \frac{1}{2}\lambda_{p4}\,h^2
  \end{pmatrix},
  \label{eq:vev:mhS}
\end{align}
while the mass of the neutral CP-odd and the charged component
of $H$ are given by
\begin{align}
  m_{G^+, G^0}^2(h,S) &=
    - \mu_H^2 + \lambda_{H4}\,h^2
    + \lambda_{p3}\,\mu_S\,S
    + \frac{1}{2} \lambda_{p4}\,S^2\,.
\end{align}
We can see that in regions where both $h$ and $S$ are non-zero, there will be
mixing between the associated particles.  Note that with our simplifying
assumption $\lambda_{p3} \ll 0$, we can neglect $h$--$S$ mixing in the $\ev{S}
= 0$, $\ev{h} \neq 0$ phase at $T=0$.
The sums in $V^\text{CW}(h,S)$ and $ V^T(h,S)$ (see \cref{eq:app:VCW,eq:app:VT})
now run over $i\in\{h,\, S,\, G^0,\, G^+,\, W^i,\, B,\, t\}$.  For $h$ and $S$,
this is understood to mean summing over the neutral CP-even mass eigenstates,
determined by diagonalising the mass matrix in \cref{eq:vev:mhS}.
The coefficients $n_i$ for the SM fields are $n_h = n_s = n_{G^0} = 1$,
$n_{G^+} = 2$, $n_{W^i} = n_B = 3$, and $n_t = -12$\,\cite{Delaunay:2007wb}.
The Debye masses of $h$ and $S$, relevant in the computation of $V^\text{daisy}$
(see \cref{eq:app:Vdaisy}) are
\begin{align}
  \Pi_{h, G^0, G^+}
    &= \frac{T^2}{48} (24\lambda_{H4} + 9g^2 + 3g'^2 + 12y_t^2 + 2\lambda_{p4})\,, \\[0.2cm]
  \Pi_S(T)
    &= \frac{T^2}{24} (\lambda_{S4} + 4\lambda_{p4} + 4y_\chi^2) \,.
\end{align}
The Debye masses of $W^i$, $B$ and $t$ are the same as in the SM (see
\cref{sec:app-eff-pot}) as these particles do not couple to $S$.  The sum in
\cref{eq:app:Vdaisy} runs over $i\in\{h, G^0,G^+,S,W^i,B\}$.  The contribution
of $\chi$ to the effective potential is negligible due to the smallness of
$y_\chi$ and is therefore dropped in our calculations.


\begin{figure}
  \begin{center}
    \includegraphics[width=0.9\textwidth]{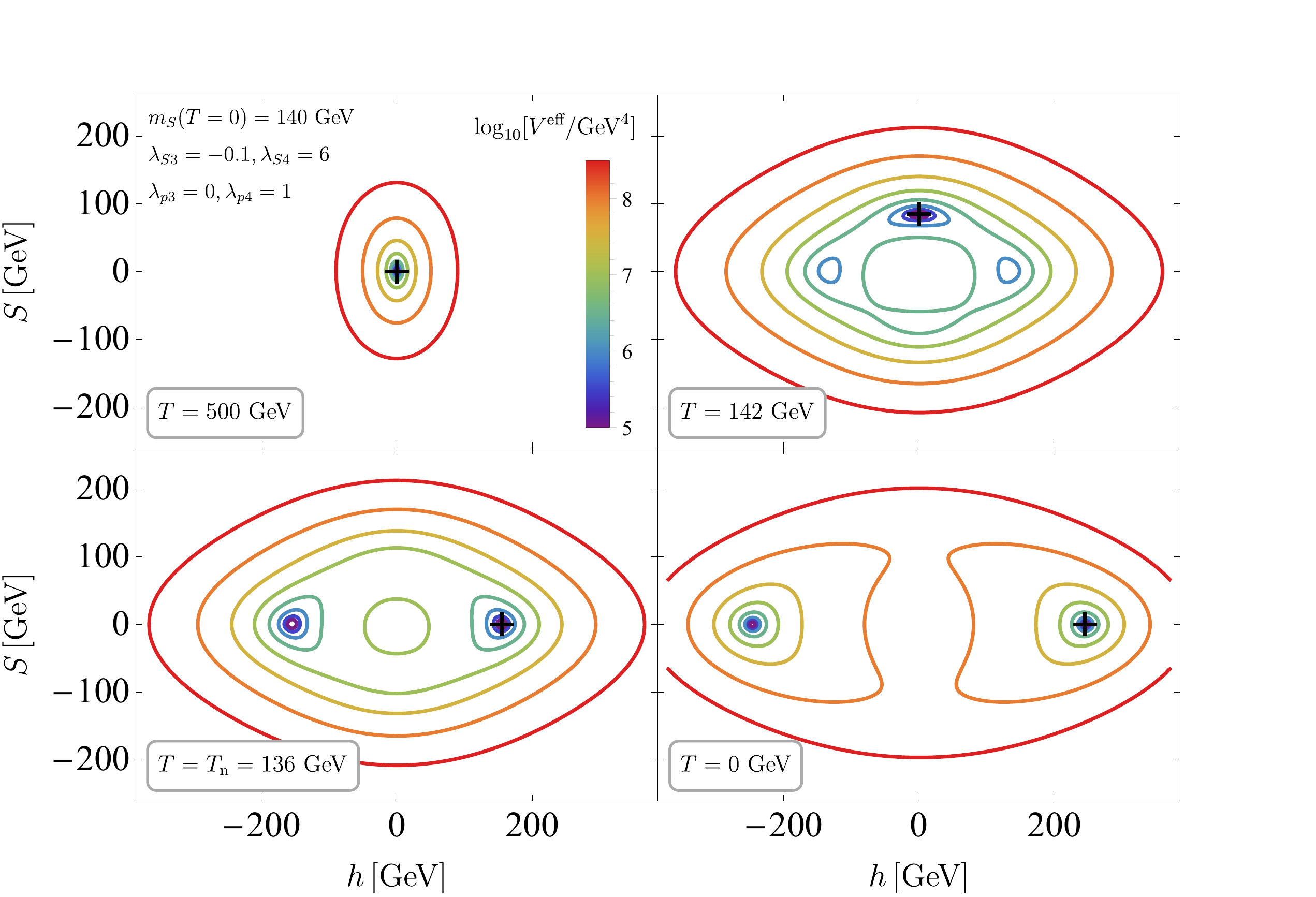}
  \end{center}
  \caption{The effective potential $V^\text{eff}$
  for a particular parameter point in the vev-induced DM production scenario
  defined in \cref{eq:vev:L}. The black cross indicates the phase the Universe is
  in at the given temperatures.  The two-step phase transition (``vev
  flip-flop'') is clearly visible:  at $T = 500\,\text{GeV}$ (top left), both
  $\ev{S}$ and $\ev{h}$ vanish, as $V^\text{eff}$ is dominated by thermal
  corrections. At a lower temperature, the Universe transitions to the $\ev{S}
  \neq 0$, $\ev{h} = 0$ phase (top right), but eventually the minimum with
  $\ev{S} = 0$, $\ev{h} \neq 0$ becomes the global one, so the Universe
  transitions into it and remains there (bottom left and bottom right).
  }
  \label{fig:vev:Veff}
\end{figure}


The behaviour of the effective potential is illustrated in \cref{fig:vev:Veff}
for a parameter point featuring a two-step phase transition or vev flip-flop.
At early times (top left panel of \cref{fig:vev:Veff}), $V^\text{eff}$
is dominated by the finite temperature and therefore approximately parabolic.
Consequently, the SM Higgs and $S$ have zero vacuum expectation values.
As the universe expands and cools down (top right panel of \cref{fig:vev:Veff}),
the finite temperature corrections become similar in magnitude to the tree-level
terms and the effective potential develops minima at $\ev{S} \ne 0$. 
There is typically a second order phase transition, so the Universe
immediately enters a phase where $S$ has a non-zero vev.  After further cooling,
new minima at $\ev{h} \ne 0$ develop.  These become the global minima at some
critical temperature $T_c$.  However, there will now be a barrier between
the $\ev{S} \ne 0$ and the $\ev{h} \ne 0$ minima, so the Universe cannot
immediately transition into the global minimum, but undergoes a short period of
supercooling.  The subsequent phase transition is first order and
proceeds via bubble nucleation, when at the nucleation temperature $T_n$ it
becomes energetically favourable for bubbles of the new phase to expand and
fill the entire universe. Typically, one finds $T_n \simeq T_c$, but in narrow regions
of parameter space, $T_n$ may also be significantly below $T_c$.
As in \cref{sec:kin}, we have used \texttt{CosmoTransitions}~\cite{Wainwright:2011kj,
Kozaczuk:2014kva, Blinov:2015sna, Kozaczuk:2015owa} to determine the 
nucleation temperature.
Numerically, \texttt{CosmoTransitions} computes $T_n$ by determining the
temperature at which $S_E(T) / T$ drops below a critical value of 140.  Here
$S_E(T)$ is the minimum Euclidean action corresponding to a transition between
the two potential minima \cite{Wainwright:2011kj}.  The effective potential at
$T = T_n$ is shown in the bottom left panel of \cref{fig:vev:Veff}.  The $h
\neq 0$ minima then deepen as $T$ goes to zero, and the universe remains in a
phase where $\ev{S} = 0$ and $\ev{h} \neq 0$.


\begin{figure}
  \begin{center}
    \includegraphics[width=0.45\textwidth]{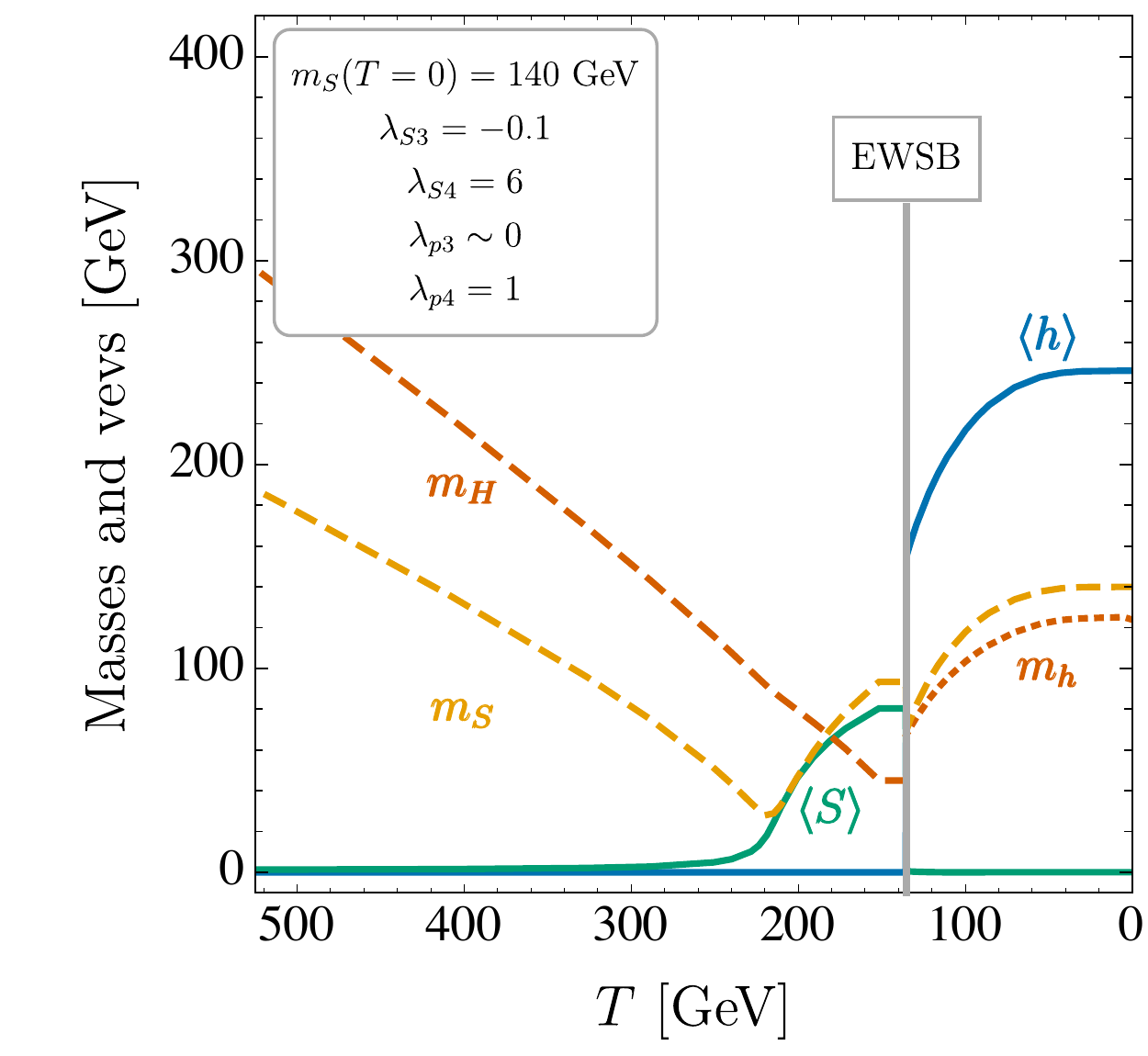}
  \end{center}
  \caption{Evolution of the scalar vevs and masses with temperature
    for a particular parameter point in the vev-induced DM freeze-in scenario.
    $m_H$ denotes the mass of the SM Higgs doublet above electroweak symmetry breaking,
    while $m_h$ is the mass of the SM-like physical Higgs boson below.}
  \label{fig:vev:evolution}
\end{figure}


In \cref{fig:vev:evolution} we show the masses and vevs of the new scalar $S$
and the SM Higgs doublet as a function of temperature.  At high temperatures
$\ev{S}$ is zero, but at lower temperatures it obtains a non-zero value.
This situation persists until the first order
electroweak phase transition at $T_n = 136\,\text{GeV}$.  At this point, $\ev{S}$
goes to zero while the SM Higgs vev becomes non-zero.  $\ev{h}$ then
gradually increases until it attains its $T=0$ value of $246\,\text{GeV}$.
We can see that, as in \cref{fig:vev:Veff}, the scalars receive large
finite temperature corrections to their masses at high temperatures.  For the
parameters chosen here, both $m_S$ and $m_h$ become smaller than their $T=0$
values between the phase transitions, similar to what we found for $m_S$
in \cref{sec:vev:Veff}.

\subsection{Dark Matter Freeze-In and Relic Abundance}
\label{sec:vev:freeze-in}

\begin{figure}
  \centering
  \includegraphics[width=0.8\textwidth]{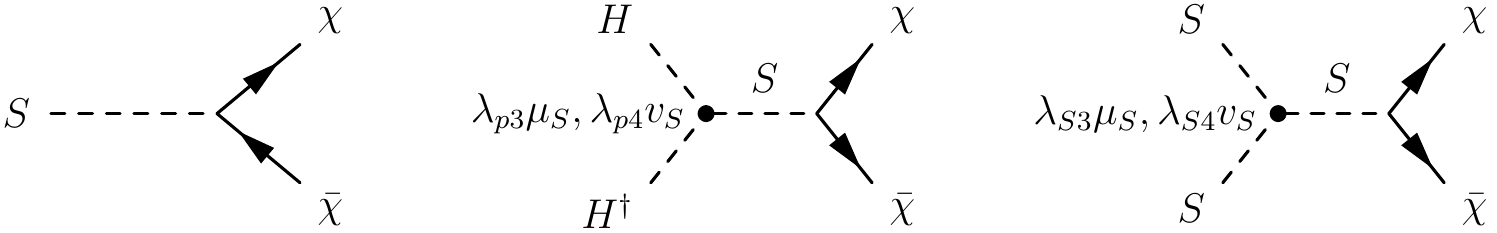}
  \caption{The main production modes for the DM particle $\chi$ in the
    vev-induced freeze-in scenario.}
  \label{fig:vev:diagrams}
\end{figure}

In the model defined in \cref{eq:vev:L}, the coupling
$\lambda_{p4}$ between the new scalar field, $S$, and the SM Higgs doublet
needs to be of order one for the vev flip-flop to occur (see discussion below
\cref{eq:vev:mS}).
Sizeable $\lambda_{p4}$ in turn means that at $T \simeq m_S,\, m_H$, the scalar $S$
is in thermal equilibrium with the SM sector. $\chi$, on the other hand,
never comes into thermal equilibrium because of our assumption that $y_\chi$
is tiny.  Instead, a small abundance of $\chi$ is produced via freeze-in,
facilitated by the processes $S \to \chi \bar\chi$, $H^\dag H \to \chi \bar\chi$,
and $S S \to \chi \bar\chi$. The corresponding Feynman diagrams are shown
in \cref{fig:vev:diagrams}, and the decay rates and annihilation cross sections are
\begin{align}
  \Gamma(S \to \chi \bar\chi) &=
    \frac{y_\chi^2}{8 \pi m_S^2(T)} (m_S^2(T) - 4 m_\chi^2)^{3/2} \,,
\\[0.2cm]
  \sigma(S S \to \chi \bar{\chi}) &=
    y_\chi^2 (\lambda_{S3} \mu_S + \lambda_{S4} v_S(T))^2
    \frac{(s - 4 m_\chi^2)^{3/2}}{8 \pi s (m_S^2(T) - s)^2 \sqrt{s - 4 m_S^2(T)}} \,,
\\[0.2cm]
  \sigma(H^\dagger H \to \chi \bar\chi) &= 
    y_\chi^2 (\lambda_{p3} \mu_S + \lambda_{p4} v_S(T))^2
    \frac{(s - 4 m_\chi^2)^{3/2}}{8\pi s (m_S^2(T) - s)^2 \sqrt{s - 4 m_h^2(T)}} \,.
\end{align}
The last expression should be understood as the cross section for one of
the two components of the doublet $H$ to annihilate with its antiparticle
into $\chi\bar\chi$.
The first process, $S \to \chi \bar\chi$, is kinematically forbidden when
$m_S(T) < 2 m_\chi$, but since thermal corrections may drive $m_S(T)$ to large
values at $T \gg m_S(T=0)$, it may be allowed at early times.  Whether or not
this production channel is important will thus depend on the reheating
temperature. We will be particularly interested in $T_R$ not too far above the
electroweak scale, as in this case the dynamics of the vev flip-flop are most
important for DM physics.  The other two freeze-in processes can be mediated
either by the cubic scalar couplings $\lambda_{S3}$, $\lambda_{p3}$, or by the
quartic couplings $\lambda_{S4}$, $\lambda_{p4}$ if $v_S \ne 0$.  We will focus
on the parameter region where the cubic couplings are small because this is the
region where freeze-in depends most strongly on $v_S$ and thus on the dynamics
of the vev flip-flop.  Moreover, as explained in \cref{sec:vev:model},
$\lambda_{S3},\, \lambda_{p3} \ll 1$ is technically natural in the 't~Hooft
sense. We note, however, that $\lambda_{p3}$ should not be exactly zero.
Otherwise, the small relic abundance of $S$ could not decay away and would
violate direct detection limits.  The small vev that $S$ has even at $T=0$ when
$\lambda_{p3} \neq 0$ would not affect our results.
It is also important to note that the
restrictions we impose here on the parameters of the scalar potential and on
$T_R$ are not necessary to make the model phenomenologically viable. There are
other large regions of parameter space where the DM relic abundance can be
successfully generated, albeit without strong involvement of the vev flip-flop.


\begin{figure}
  \begin{center}
    \begin{tabular}{cc}
      \includegraphics[width=0.45\textwidth]{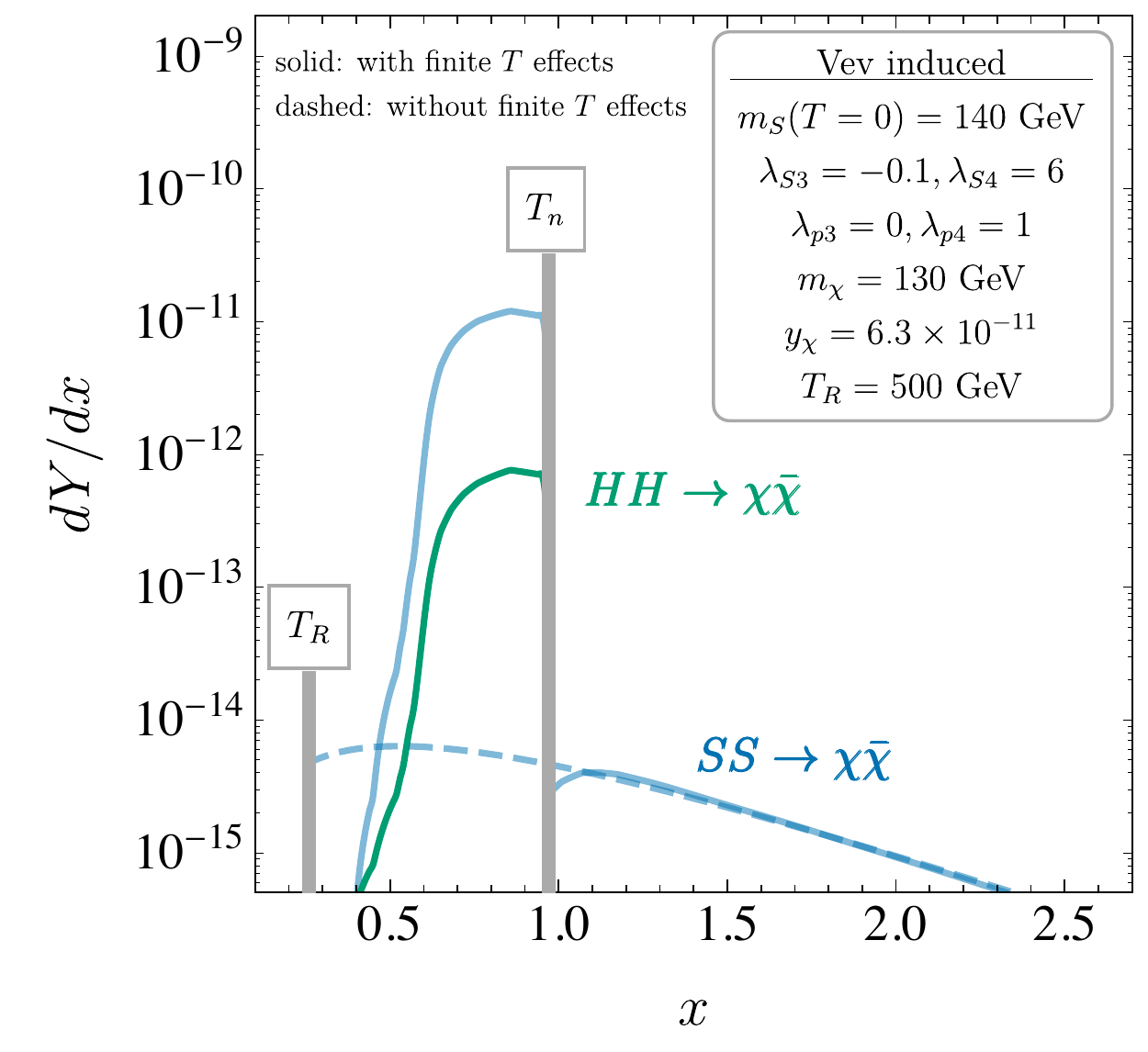} &
      \includegraphics[width=0.45\textwidth]{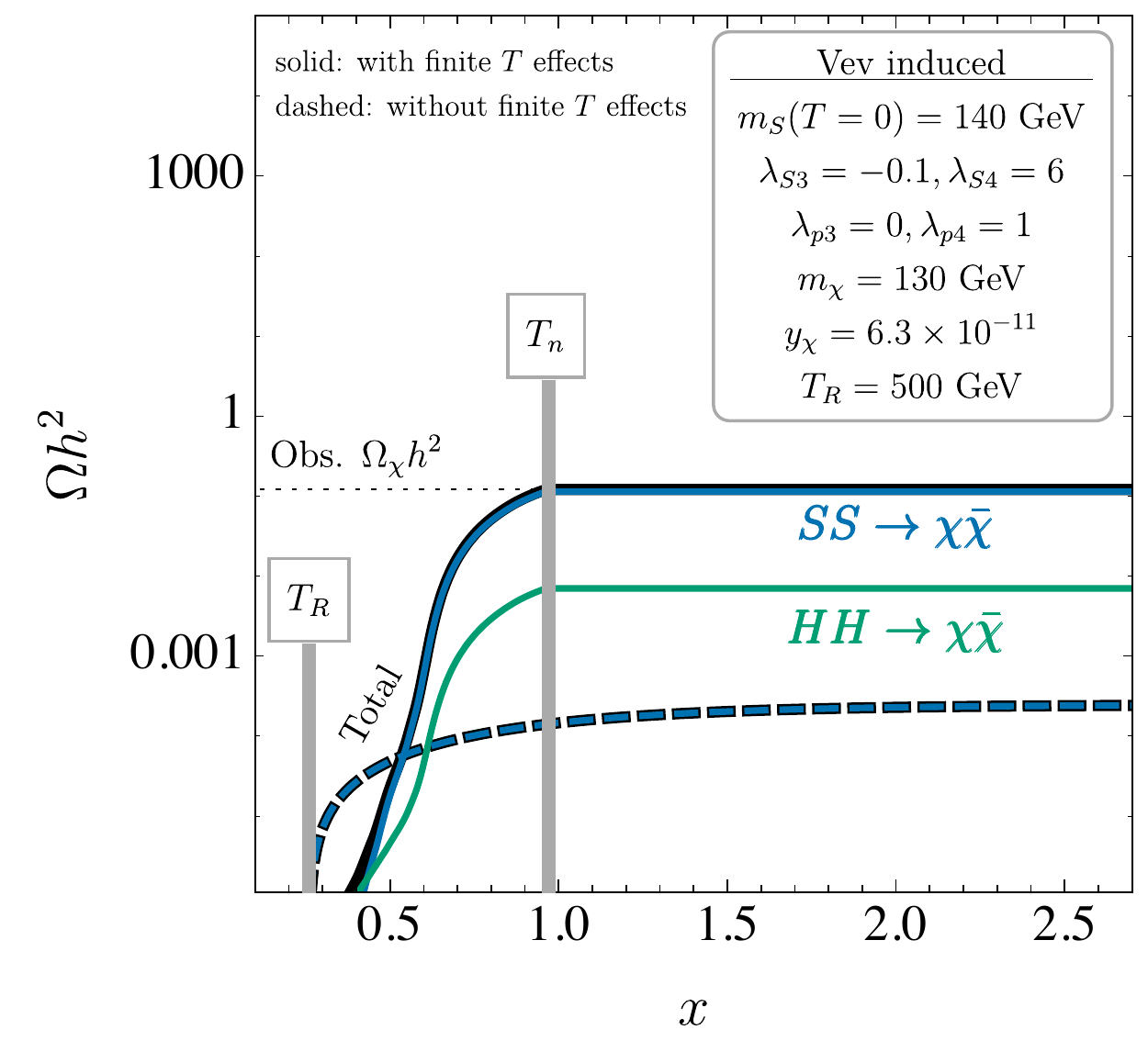} \\
      \includegraphics[width=0.45\textwidth]{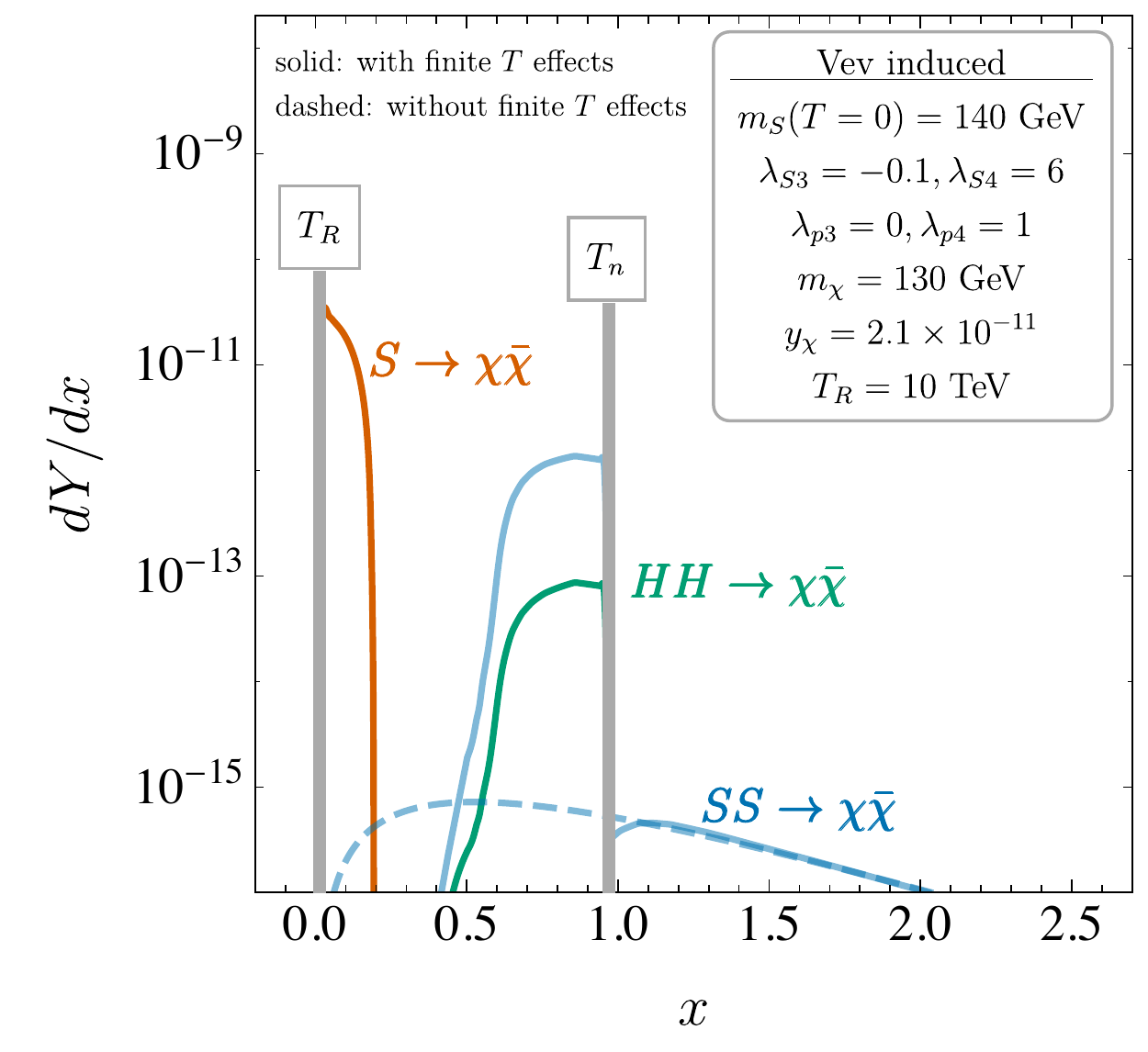} &
      \includegraphics[width=0.45\textwidth]{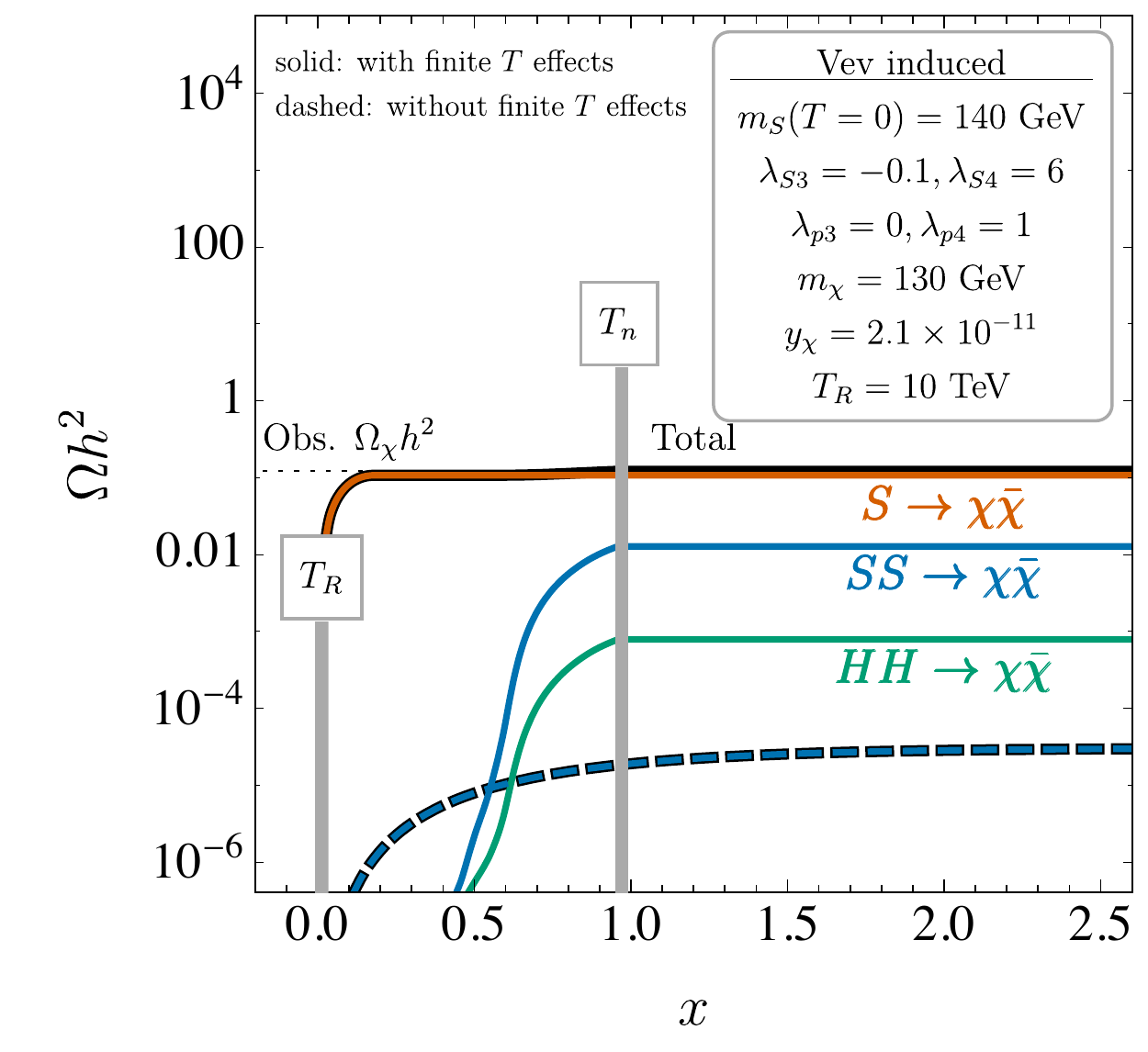} \\
    \end{tabular}
  \end{center}
  \caption{
    Evolution of the DM production rate (left) and the DM abundance
    extrapolated to $z=0$ (right) as a function of $x \equiv m_\chi / T$ 
    for two different parameter points in the
    vev-induced freeze-in scenario defined by \cref{eq:vev:L}.
    Solid curves correspond to the different production mechanisms shown in
    \cref{fig:vev:diagrams}. The dashed blue line indicates the
    result one would obtain if thermal corrections and the vev flip-flop were
    neglected. We see that for a reheating temperature $T_R$ just above the
    electroweak scale (top panels), the DM abundance is entirely dominated by
    the processes $H^\dag H \to \chi \bar\chi$ and $S S \to \chi \bar\chi$,
    whose rates are greatly enhanced when $\ev{S} \neq 0$.
     For larger $T_R$ (bottom
    panels), the process $S \to \chi \bar\chi$, which is independent of
    $\ev{S}$, becomes allowed thanks to thermal corrections, even though at
    $T=0$, $m_S < 2 m_\chi$.}
  \label{fig:vev:dYdx}
\end{figure}


The DM production rate and the resulting abundance (extrapolated to
zero redshift) are shown in \cref{fig:vev:dYdx} for two illustrative
parameter points of the vev-induced freeze-in scenario.
The first parameter point shown in \cref{fig:vev:dYdx} (top panels) is
characterised by a low reheating temperature $T_R = 500$\,GeV.  In this case,
DM production is entirely dominated by $2 \to 2$ processes proportional to
$v_S$, so the dynamics of the vev flip-flop are crucial in this case.  We
see that the DM production rate $dY/dx$ rises rapidly after $S$ develops a vev
at $T \simeq250\,\text{GeV}$. Between the two phase transitions, $dY/dx$ follows
the evolution of $v_S$, and the contribution from the $v_S$-dependent
channels is 2--3 orders of magnitude larger than the contribution from
vev-independent $S S \to \chi \bar\chi$ annihilation via $\lambda_{S3}$.  Among
the vev-dependent channels, $S S\to \chi \bar\chi$ dominates over 
$H^\dag H \to \chi \bar\chi$ mainly because $\lambda_{S4} > \lambda_{p4}$.
The small drop in $dY/dx$ immediately before the electroweak phase transition 
is due to the onset of Boltzmann suppression of $H$ and $S$. 
After the electroweak phase transition at $T \simeq136\,\text{GeV}$, the
$v_S$-dependent production processes cease. Before and after the two phase
transitions, only the $v_S$-independent channel $S S \to \chi \bar\chi$ via
$\lambda_{S3}$ is active, but its overall contribution is tiny.  At $x \lesssim 0.5$, 
this channel is suppressed as the $s$-channel mediator, 
$S$, is very heavy at these high temperatures.  Beyond the 
electroweak phase transition this channel gradually reduces, due to the 
Boltzmann suppression of $S$.

At the second parameter point shown in \cref{fig:vev:dYdx} (bottom panels)
the behaviour of the vev-dependent DM production channels and of vev-independent
production via $S S \to \chi \bar\chi$ (mediated by $\lambda_3$) is similar to
the top panels.  However, as the second parameter point features a larger
$T_R = 10$\,TeV, the decay channel $S \to \chi \bar\chi$, which does not
depend on $v_S$ and on the vev flip-flop, is kinematically allowed at $x \lesssim 0.2$,
when thermal corrections lift $m_S$ above $2 m_\chi$.  In this case,
this production channel dominates the final abundance.

We conclude that, for low $T_R$, it is the dynamics of the vev flip-flop
that determines the DM abundance today. For high $T_R$, it is the
thermal corrections to $m_S(T)$, which in turn depend on the couplings in
the scalar sector, especially $\lambda_{S4}$.  In either case, the inclusion
of thermal effects in the computation of the DM relic density is
essential.  To emphasise this point, we show in \cref{fig:vev:dYdx}
also the production rate and abundance that would be obtained if thermal
corrections to $V^\text{eff}$ (and thus the two-step phase transition) were
neglected in the calculation (dashed blue lines).  In this case, only the
processes $S S \to \chi \bar\chi$ and $H^\dag H \to \chi \bar\chi$ with cross
sections proportional to the small couplings $\lambda_{S3}^2$ and
$\lambda_{p3}^2$, respectively, would contribute.  For our choice $\lambda_{p3}
= 0$, only $S S \to \chi \bar\chi$ is open.  The production
rate in this channel is non-zero at $T_R$ (or even during
preheating~\cite{Garcia:2017tuj}, which we neglect here assuming it is very
rapid) and first rises slightly as there is more time to freeze-in at greater $x$.  
At $x \gtrsim 0.5$, the rate
begins to drop as Boltzmann suppression becomes significant.
Eventually, $\lambda_{S3}$-mediated production via $S S \to \chi
\bar\chi$ freezes out.


\begin{figure}
  \begin{center}
      \includegraphics[width=0.45\textwidth]{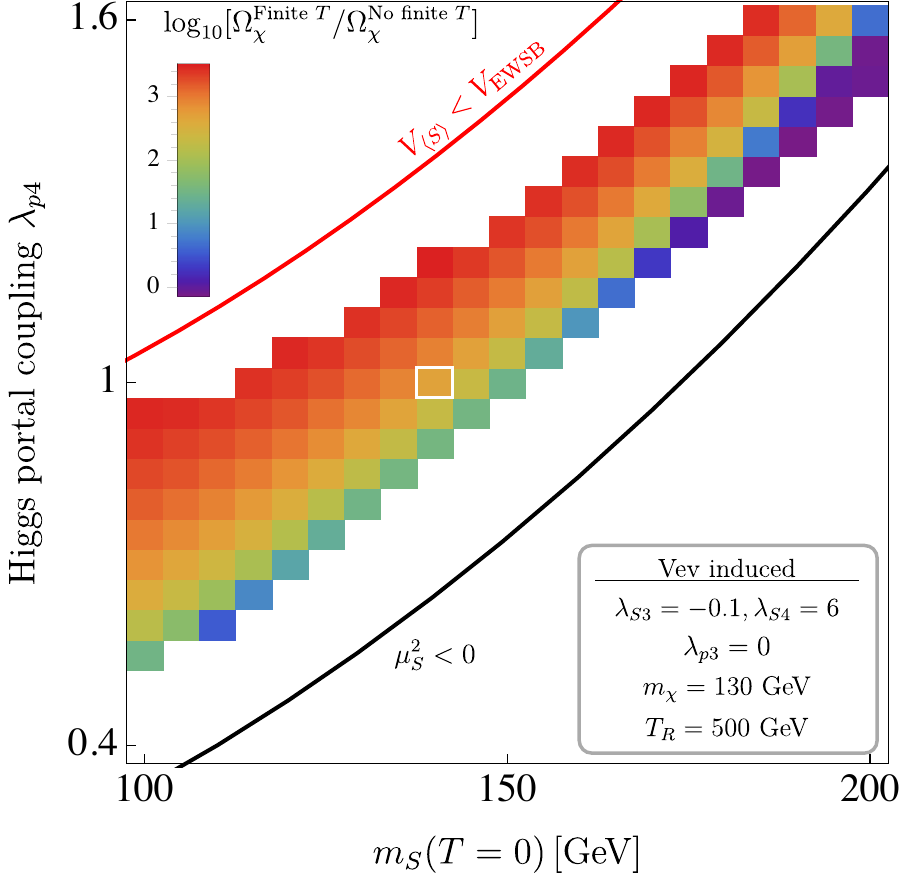}
  \end{center}
  \caption{
    The ratio of the DM abundance including finite temperature effects to that
     ignoring finite
    temperature effects in the vev-induced freeze-in scenario.
    We show a cut through the parameter space of the
    model in the plane spanned by the zero temperature
    mass of $S$, $m_S(T=0)$, and the quartic Higgs portal coupling
    $\lambda_{p4}$.   The white outline indicates the 
    point considered in \cref{fig:vev:dYdx}.}
  \label{fig:vev:paramspace}
\end{figure}


We further explore the crucial importance of thermal corrections in
\cref{fig:vev:paramspace}, which shows a cut through the model's parameter
space in the plane spanned by the zero temperature mass of $S$, $m_S(T=0)$, and
the quartic Higgs portal coupling $\lambda_{p4}$.  The pixelated region 
shows where the two-step phase transition occurs.  The colour coding quantifies
the ratio of the DM abundance obtained including thermal corrections to the
abundance if these corrections were neglected for a low $T_R$.  We see that
thermal effects dominate the abundance by up to four orders of magnitude.  For fixed $m_\chi$,
they are largest at small $m_S(T=0)$, where thermal corrections to $m_S$ are most
important, and at large $\lambda_{p4}$, where the $\ev{S} \neq 0$ phase
lasts longer.  
For a high $T_R \sim 10 \,\text{TeV}$, the freeze-in abundance is dominated by the 
$S\to\chi\bar{\chi}$ channel and thermal effects dominate the abundance by around 
four orders of magnitude over the whole parameter space shown.
In the white area at the top of \cref{fig:vev:paramspace}, the
global minimum of $V^\text{eff}$ at $T=0$ would be the one with $\ev{S}\neq 0$, i.e.,
electroweak symmetry would never be broken. In the white region at the bottom
of the plot, $S$ never acquires a non-zero vev.

\section{Vev-Induced Mixing with a Vev Flip-Flop}
\label{sec:mix}

\begin{figure}
  \centering
  \includegraphics[width=0.5\textwidth]{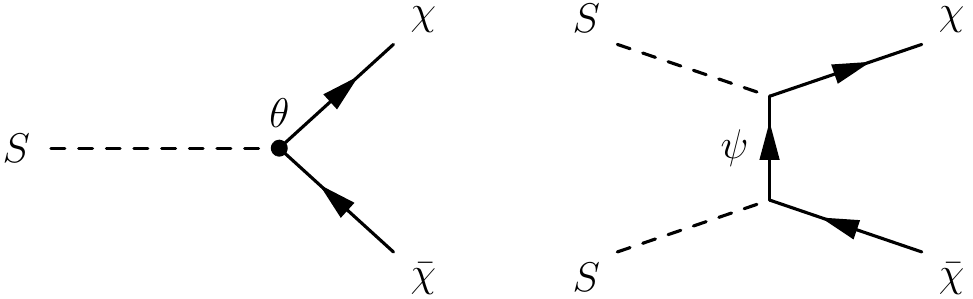}
  \caption{Feynman diagrams contributing to DM freeze-in in the
    mixing induced scenario. $\theta$ denotes the $\ev{S}$-dependent mixing
    angle between $\chi$ and $\psi$.}
  \label{fig:mix:diagrams}
\end{figure}

Let us now move to a third scenario illustrating the importance of thermal
effects on the DM abundance in the Universe.  The scenario discussed in the
following, which we dub mixing induced freeze-in, is based on the same particle
content as the kinematically induced freeze-in model from \cref{sec:kin} with an 
extra discrete symmetry.  The
model's Lagrangian is thus given by \cref{eq:kin:L,eq:kin:V}, with the two forbidden Yukawa 
terms removed. The most
important term for DM freeze-in is again the Yukawa coupling $y_{\psi\chi}
\bar\psi S \chi$.  However, we now assume the DM candidate $\chi$ and the new
scalar $S$ to have masses around the electroweak scale, with $m_S > m_\chi$.
The auxiliary new fermion $\psi$ is assumed to be much heavier (see
\cref{tab:mix:particles}).  The idea is that the reheating temperature is low,
$T_R < m_\psi$, so that $\psi$ never comes into thermal equilibrium and DM
production via $\psi \to S \chi$ (the channel we had focused on in
\cref{sec:kin}) does not occur.  Instead, the main DM production channels will
be $S \to \chi \bar\chi$, facilitated by $\chi$--$\psi$ mixing through the $S
\bar\chi \psi$ coupling, and $S S \to \chi \bar\chi$, mediated by a $t$-channel
$\psi$ (see \cref{fig:mix:diagrams}).  The former process is of particular
interest to us because $\chi$--$\psi$ mixing depends on the vev of $S$.  For
the parameters of the scalar potential, we consider values similar to the ones
we chose (and motivated) in \cref{sec:vev}: negligible $\lambda_{S3}$,
$\lambda_{p3}$, but sizeable Higgs portal and dark sector quartic couplings,
$\lambda_{p4},\, \lambda_{S4} \simeq \mathcal{O}(1)$, to induce a vev flip-flop.
Thus, DM production via $S \to \chi \bar\chi$ will be open for a limited amount
of time while $v_S \neq 0$.  In the following, we will study the interplay of
the two production processes, focusing in particular on the importance of $v_S$
and the vev flip-flop.


\begin{table}
  \begin{center}
  \begin{minipage}{11cm}
    \begin{ruledtabular}
    \begin{tabular}{ccccc}
      Field  &   Spin        & $\mathbb{Z}_2$ & $\mathbb{Z}_2'$ & mass scale \\
      \hline
      $S$    &    $0$        &    $+1$        & $-1$            & $m_S(T=0) \simeq 100\,\text{GeV}$ \\
      $\chi$ & $\frac{1}{2}$ &    $-1$        & $-1$            & $m_\chi \simeq 100\,\text{GeV}$ \\
      $\psi$ & $\frac{1}{2}$ &    $-1$        & $+1$            & $m_\psi> 10^3\,\text{GeV}$ \\
      \bottomrule
    \end{tabular} 
    \end{ruledtabular}
  \end{minipage}
  \end{center}
  \caption{The new particle content in the mixing induced scenario.
    All fields are SM singlets.}
  \label{tab:mix:particles}
\end{table}


The decay width for $S \to \chi \bar\chi$ is
\begin{align}
  \Gamma(S \to \chi \bar\chi) &= 
    \frac{(y_{\psi\chi} \sin\theta_{\psi\chi})^2}{8 \pi m_S^2(T)}
    (m_S^2(T) - 4m_\chi^2)^{3/2} \,,
  \label{eq:mix:Gamma-S-chi-chi}
\intertext{and the cross section for $S S \to \chi \bar\chi$ reads}
  \sigma(S S \to \chi \bar\chi) &\approx 
    \frac{y_{\psi\chi}^4}{8\pi m_\psi^2} 
    \frac{(s - 4 m_\chi^2)^{3/2}}{s \sqrt{s - 4m_S^2(T)}} \,,
  \label{eq:mix:sigma-S-S-chi-chi}
\end{align}
where we have taken the limit $m_\psi \gg \sqrt{s}, m_S(T), m_\chi$, 
which is justified in view of our assumption $m_\psi \gg T_R$.  
The mixing angle between $\chi$ and $\psi$ is
\begin{align}
  \theta_{\psi\chi} \simeq \frac{y_{\psi\chi} v_S(T)}{m_\psi} \,.
  \label{eq:mix:theta}
\end{align}


\begin{figure}
  \begin{center}
    \begin{tabular}{cc}
      \includegraphics[width=0.45\textwidth]{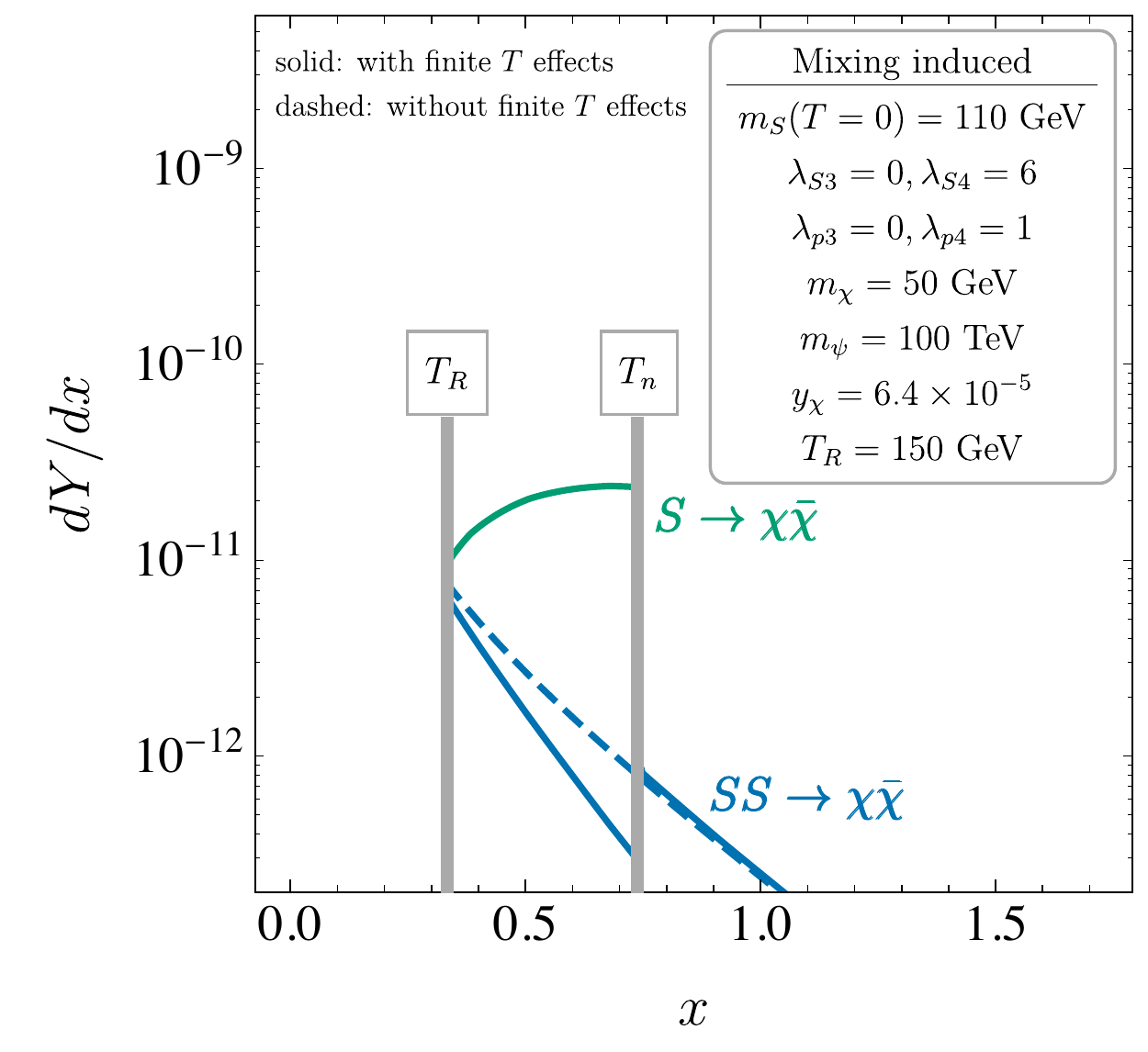} &
      \includegraphics[width=0.45\textwidth]{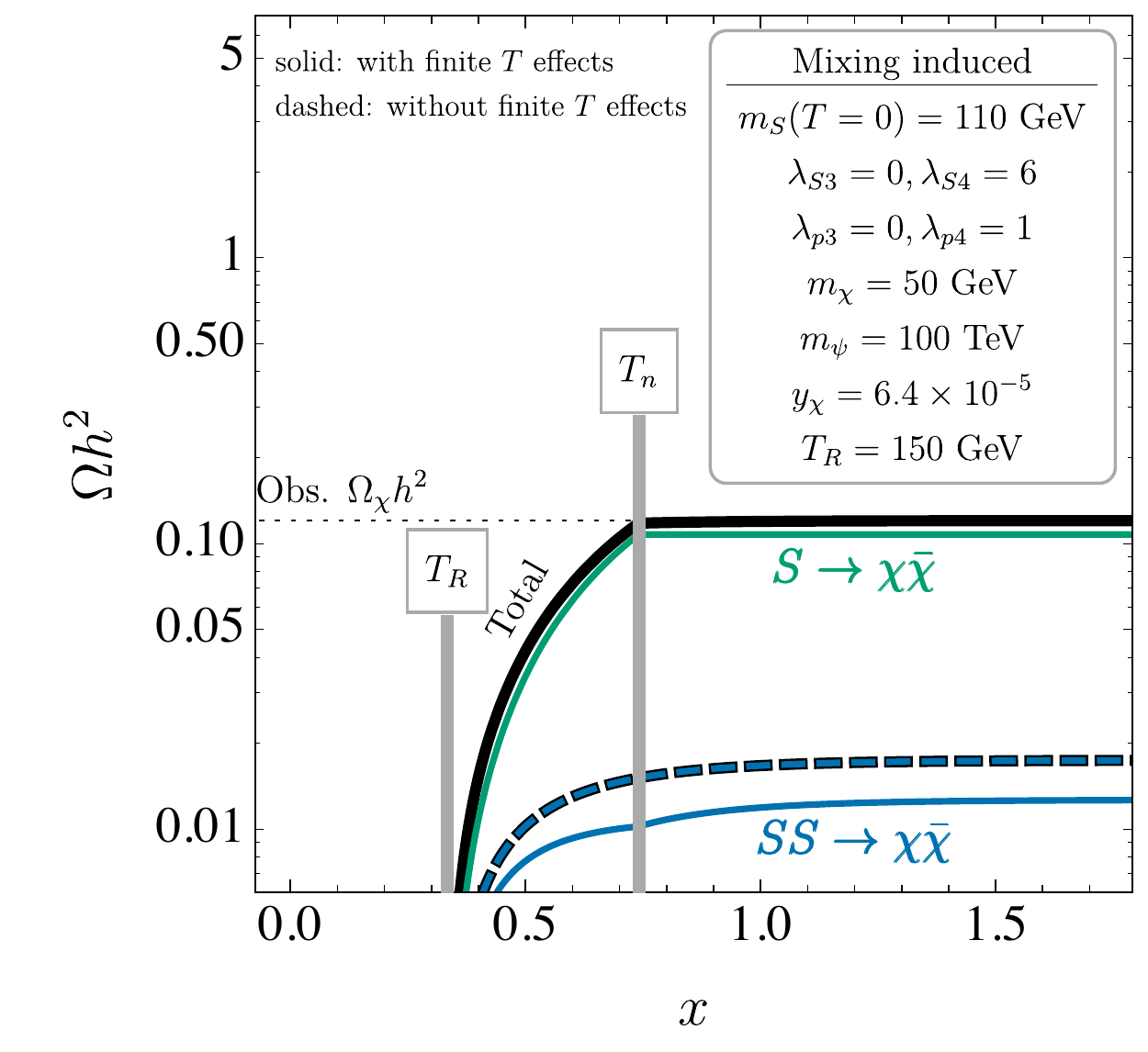} \\
      \includegraphics[width=0.45\textwidth]{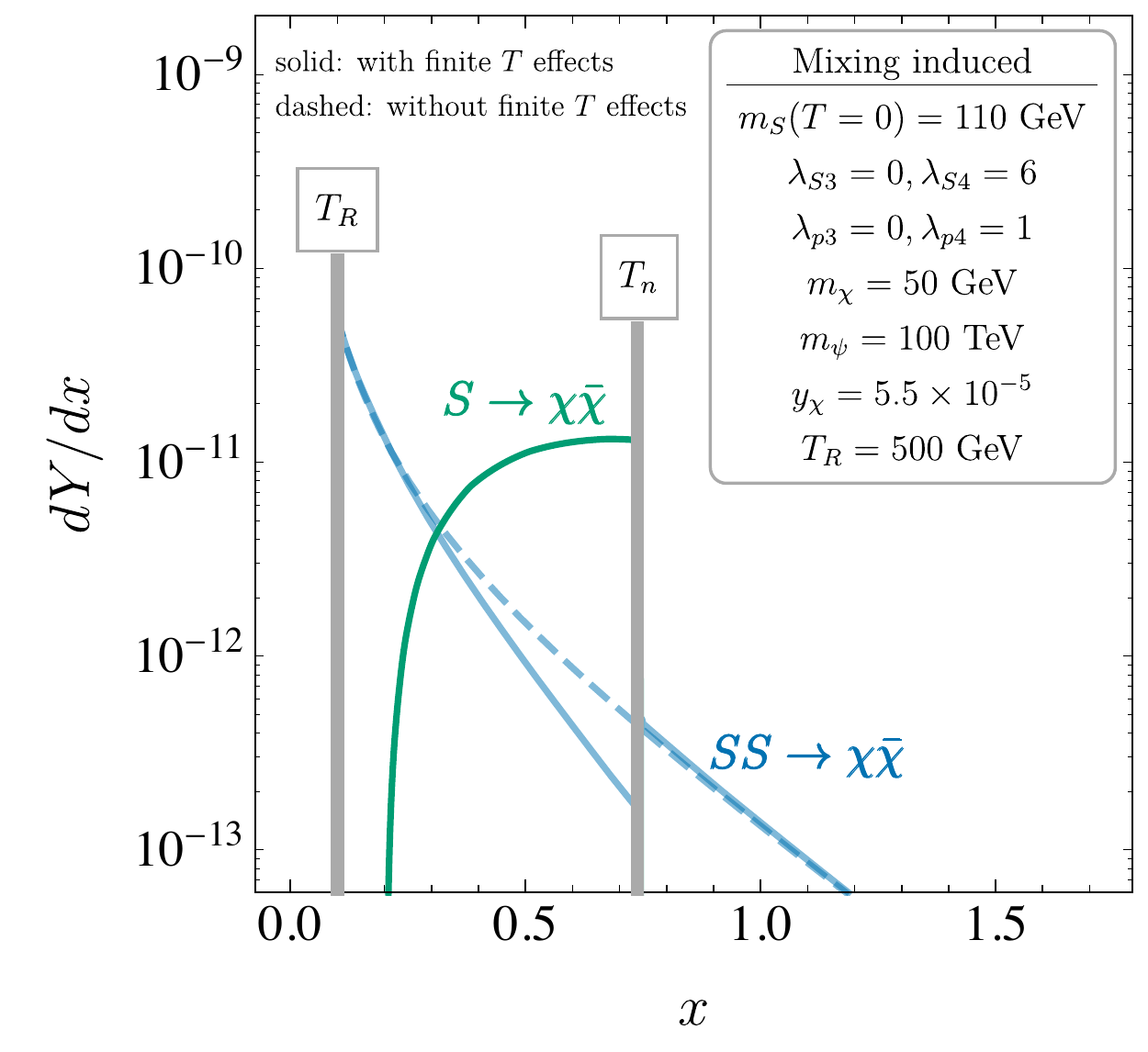} &
      \includegraphics[width=0.45\textwidth]{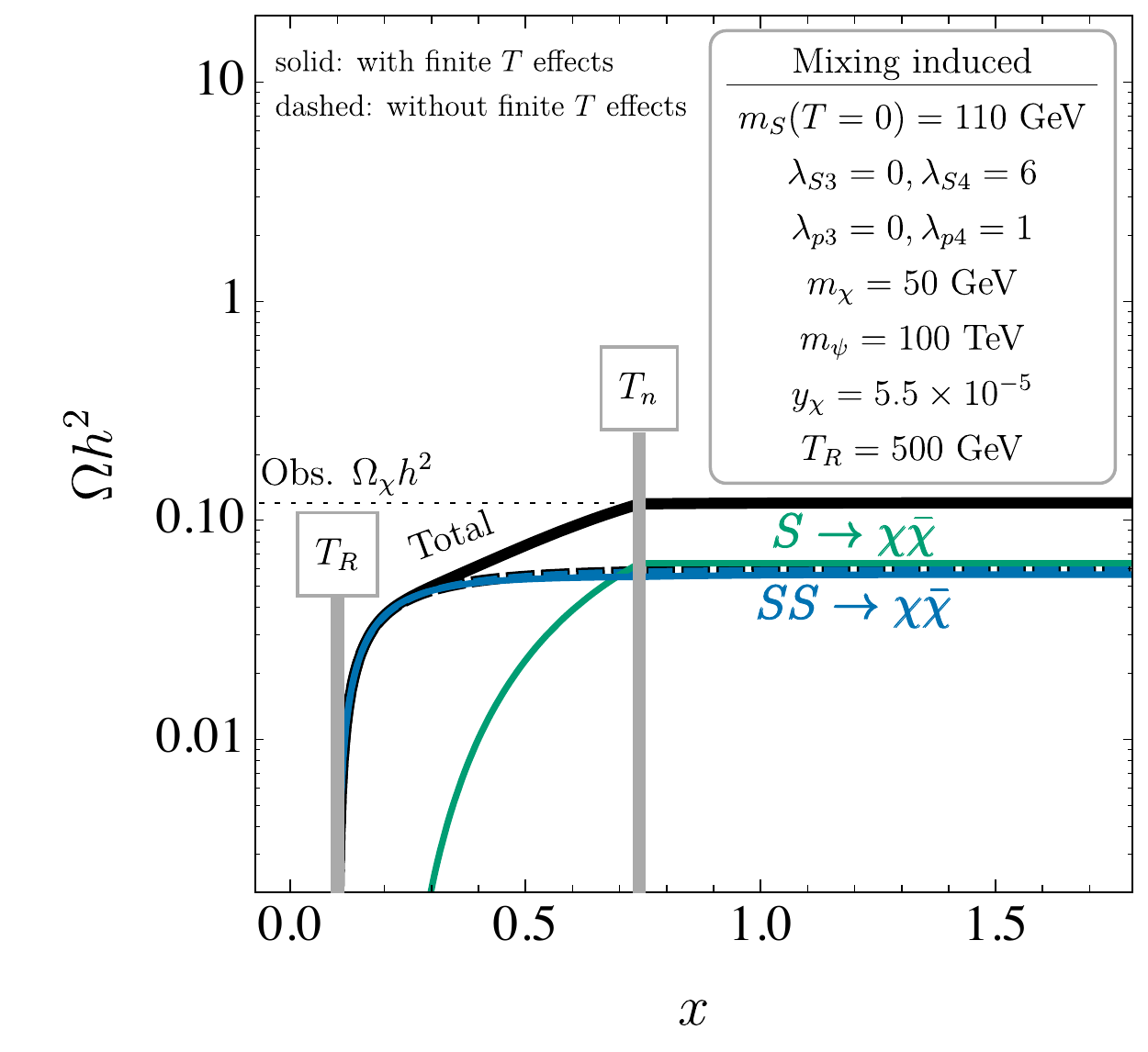} \\
    \end{tabular}
  \end{center}
  \caption{Evolution of the DM production rate (left) and the DM abundance
    extrapolated to $z=0$ (right) as a function of $x \equiv m_\chi / T$ for two different parameter points in the
    mixing induced scenario from \cref{sec:mix}.
    Solid curves correspond to the different production mechanisms shown in
    \cref{fig:mix:diagrams}. The dashed curves indicates the
    result one would obtain if thermal corrections and the vev flip-flop were
    neglected. We see that for a reheating temperature $T_R$ just above the
    electroweak scale (top panels), the DM abundance is dominated by
    the mixing-induced decay $S \to \chi \bar\chi$,
    which is facilitated by the vev flip-flop.  For larger $T_R$ (bottom
    panels), the process $S S \to \chi \bar\chi$, which is independent of
    $\ev{S}$, becomes relevant.}
  \label{fig:mix:dYdx}
\end{figure}


The dynamics of freeze-in via mixing are illustrated in
\cref{fig:mix:dYdx}. In analogy to \cref{fig:vev:dYdx} in \cref{sec:vev}, this
figure shows the DM production rate and the extrapolated DM abundance as a
function of $x \equiv m_\chi / T$.  We again consider one parameter point with
a low reheating temperature, $T_R = 150$\,GeV (top panels), and one parameter
point with a higher reheating temperature, $T_R = 500$\,GeV (bottom panels).
For low $T_R$, the DM abundance today is dominated by the decay $S \to \chi
\bar\chi$, which is only possible for $v_S \neq 0$.  For the parameters shown
in the top panels of \cref{fig:mix:dYdx}, this phase is already realised at
$T_R$.  The rate for $S \to \chi \bar\chi$ increases along with $v_S(T)$ and
then drops sharply to zero at $x \simeq 0.7$ as the electroweak phase
transition switches $v_S$ off in favour of non-zero $v_H$.  Similar behaviour is
also seen for higher reheating temperature (bottom panels of
\cref{fig:mix:dYdx}), but in this case, the overall importance of $S \to \chi
\bar\chi$ does not dominate that of $S S \to \chi \bar\chi$. As the latter process
is independent of $v_S$, it leads to DM production immediately after reheating
(or already during preheating, which we neglect here), while $S \to \chi
\bar\chi$ is only activated when $v_S$ becomes non-zero at $x \simeq 0.2$.
Note from \cref{eq:mix:sigma-S-S-chi-chi} that for $T_R < m_\psi$ (the case
realised in \cref{fig:mix:dYdx}), the DM production rate via $S S \to \chi
\bar\chi$ scales $\propto T^3 / m_\psi^2$. In other words, freeze-in is
ultraviolet-dominated~\cite{Elahi:2014fsa}.  The dependence of $S S \to \chi
\bar\chi$ on thermal corrections (namely the temperature dependence of $m_S(T)$)
is relatively weak. It is reflected for instance in a jump in the rate at
the electroweak phase transition.  For comparison, we show also the production
rate and DM yield in the absence of thermal corrections (dashed lines).
We see that a calculation neglecting these corrections would fairly accurately predict
freeze-in via $S S \to \chi \bar\chi$, but would completely miss the
mixing-induced channel $S \to \chi \bar\chi$.


\begin{figure}
  \begin{center}
    \begin{tabular}{cc}
      \includegraphics[width=0.45\textwidth]{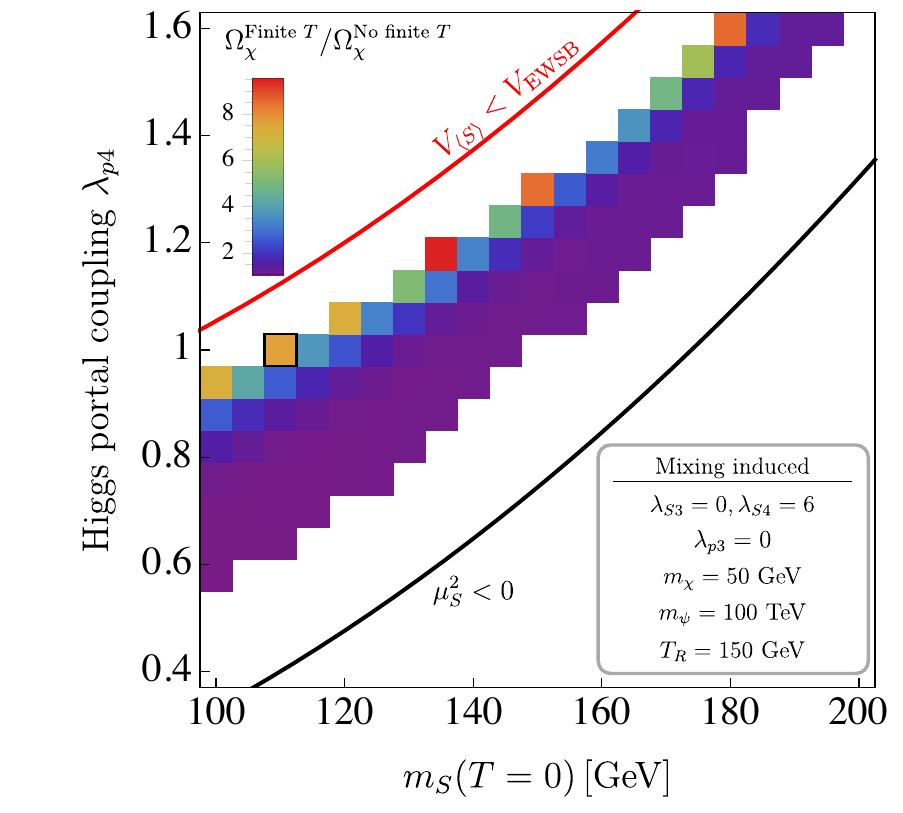} &
      \includegraphics[width=0.45\textwidth]{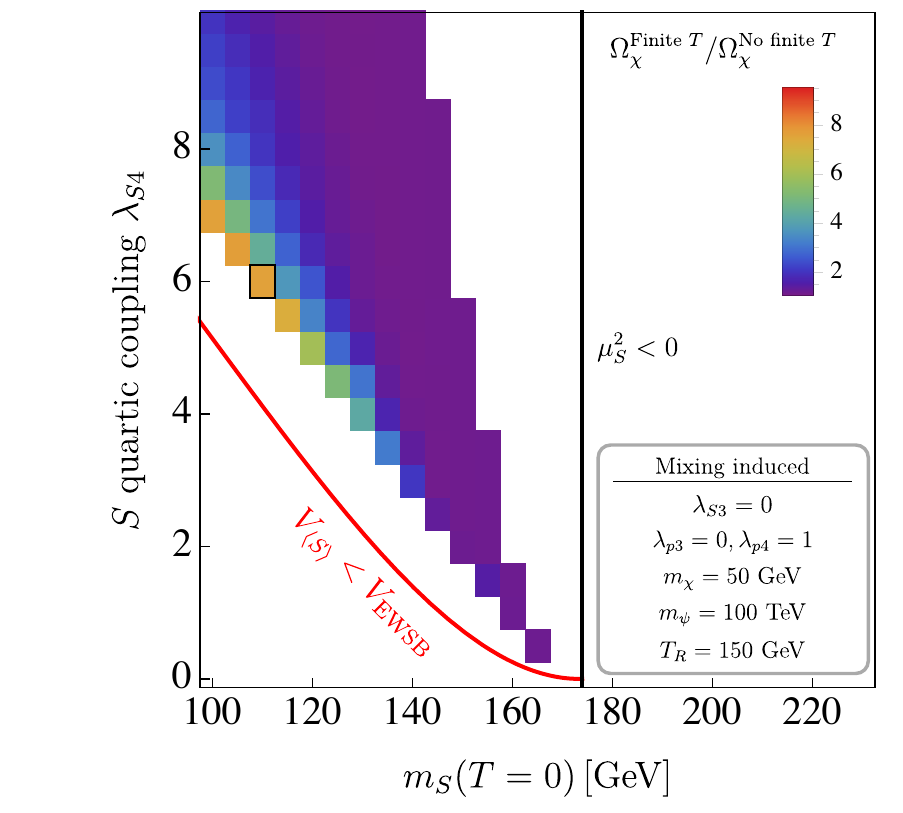} \\
    \end{tabular}
  \end{center}
  \caption{The ratio of the DM abundance including over ignoring finite
    temperature effects in the mixing induced scenario. 
    We show a cut through the parameter space in the plane spanned by
    the zero temperature mass of $S$, $m_S(T=0)$, and the quartic Higgs portal
    coupling $\lambda_{p4}$ (left) and a slice of parameter space in the $m_S(T=0)$--$\lambda_{S4}$
    plane (right). We see that the impact of thermal effects is largest at large
    portal coupling $\lambda_{p4}$ and small self-coupling $\lambda_{S4}$,
    where the $\ev{S} \neq 0$ phase lasts longer.  The black outline indicates the 
    point considered in \cref{fig:mix:dYdx}.
  }
  \label{fig:mix:paramspace}
\end{figure}


We further explore the dependence of thermal effects on the parameters of the 
mixing induced scenario in \cref{fig:mix:paramspace}.  The two panels in
this figure show different slices through the parameter space: one in the
$m_S(T=0)$--$\lambda_{p4}$ plane (cf.\ \cref{fig:vev:paramspace}) and one in the
$m_S(T=0)$--$\lambda_{S4}$ plane. We observe that thermal effects --- in this case
the $v_S$-dependent process $S \to \chi \bar\chi$ --- are most relevant both when
the Higgs portal coupling $\lambda_{p4}$ is large and when the quartic self-coupling
$\lambda_{S4}$ or the $T=0$ $S$ mass is small.  For larger $\lambda_{p4}$ or smaller $m_S(T=0)$, the
$v_S\neq 0$ vacuum becomes deeper and the phase during which $S$ has a vev
and $S \to \chi \bar\chi$ is open becomes longer.  For small $\lambda_{S4}$, the
$v_S \neq 0$ phase begins earlier, again implying that there is more time
for DM production via mixing.  In any case, we see that thermal effects are at
most $\mathcal{O}(1)$ in most of the parameter space, but in some regions can
modify the DM abundance today by an order of magnitude.

\section{Summary and Conclusions}
\label{sec:conclusions}

In this work we have considered the impact of finite temperature corrections 
on the freeze-in of dark matter.  We have highlighted several effects which 
can have a dramatic impact on dark matter production.  We have illustrated 
the impact of these effects in three toy models, which demonstrate `kinematically 
induced', `vev induced' and `mixing induced' freeze-in, respectively.

In `kinematically induced freeze-in', the dominant production channel of dark 
matter may be closed at zero temperature, but may be open in the early universe 
as temperature-dependent particle masses vary.  Although calculationally complex, 
a simple realisation of this is the SM Higgs coupling to two dark sector fermions.  We 
have highlighted the analogous effect in a realistic toy model consisting of a new scalar 
which is weakly coupled to the SM, and two dark sector fermions.  We show that 
a calculation ignoring the temperature-dependent scalar mass produces 
an estimate of the dark matter abundance which is incorrect by a factor of $\mathcal{O}(10)$.  

If instead, the new scalar couples significantly to the SM Higgs, the Higgs portal coupling 
can induce a two-step phase transition (or ``vev flip-flop'').  In this case the new scalar 
may obtain a vev for some time, which then disappears when the SM Higgs obtains its 
vev.  This can lead to `vev induced' production, where dominant channels of dark matter 
production only open when the new scalar has a vev.  We illustrate the effect in a phenomenologically 
viable toy 
model.  For reheating temperatures around the electroweak scale, 
dark matter production occurs mainly via `vev induced' production, whereas for higher 
reheating temperatures ($T_R \gtrsim$ a few TeV) `kinematically induced freeze-in' 
dominates the production.  In both cases, these finite temperature effects
 can easily change the relic abundance by 
several orders of magnitude.

Finally we consider a scenario where the temporary vev of a new scalar leads to 
mixing between fermionic dark matter and another fermion, which we call 
`mixing induced freeze-in'.  This mixing may then offer a dark matter production mode 
which is not available at zero-temperature.  We again consider a realistic toy model and 
show that dark matter can be dominantly produced through this temperature-dependent 
channel.  The error introduced by ignoring this effect can be as large as a factor of $\sim 10$, 
but depends crucially on a reheating temperature around the electroweak scale and a long 
period of vev induced mixing.  
When the reheating temperature is much higher or the vev induced mixing is weak, 
the standard calculation provides 
a reliable estimate.

\section*{Acknowledgments}

It is a pleasure to thank Christophe Grojean, Matthias K\"onig, and Andrea
Thamm for useful discussions. MJB would like to thank CERN for warm hospitality
during part of this work.  This work has been funded by the German Research
Foundation (DFG) under Grant Nos.\ EXC-1098, \mbox{KO~4820/1--1}, FOR~2239,
GRK~1581, and by the European Research Council (ERC) under the European Union's
Horizon 2020 research and innovation programme (grant agreement No.\ 637506,
``$\nu$Directions'').  MJB was also supported by the Swiss National Science
Foundation (SNF) under contract 200021-175940.

\appendix
\label{sec:app}

\section{Computation of the Effective Potential}
\label{sec:app-eff-pot}

In this appendix, we review the computation of the effective potential
$V^\text{eff}(h,S,T)$ at non-zero temperature.  The leading terms in
$V^\text{eff}(h,S,T)$ are the temperature-independent tree level potential
$V^\text{tree}(h,S)$, the Coleman Weinberg correction $V^\text{CW}(h,S)$~\cite{Coleman:1973jx}, 
and a counterterm $V^\text{CT}(h,S)$, as
well as the temperature dependent 1-loop thermal corrections $V^T(h,S,T)$~\cite{Dolan:1973qd}
and a contribution from resummed higher order ``daisy'' diagrams~\cite{Carrington:1991hz,
Quiros:1999jp, Delaunay:2007wb}:
\begin{align}
  V^\text{eff}(h,S,T) &\simeq V^\text{tree}(h,S)  +  V^\text{CW}(h,S)
                       + V^\text{CT}(h,S)
                       + V^T(h,S,T) + V^\text{daisy}(h,S,T)\,.
  \label{eq:app:Veff}
\end{align}
The tree level potential $V^\text{tree}$ is simply read off from the
Lagrangian.  For the models considered in this paper, it is given by
\cref{eq:kin:V}.  The $T$-independent Coleman-Weinberg contribution
is~\cite{Coleman:1973jx, Quiros:1999jp}
\begin{align}
  V^\text{CW}(h,S) &=
    \sum_i \frac{n_i}{64 \pi^2} m_i^4(h,S)
    \bigg[ \log\bigg( \frac{m_i^2(h,S)}{\Lambda^2} \bigg)
               - C_i \bigg] + V^\text{CT}(h,S) \,,
  \label{eq:app:VCW}
\end{align}
where the sum is over the eigenvalues of the mass matrices of all fields which
couple to the scalars, and $|n_i|$ accounts for their respective numbers of
degrees of freedom. $n_i$ is positive for bosons and negative for fermions.  We
take the renormalisation scale $\Lambda$ to be the SM Higgs vev $v_H =
246$\,GeV.  In the dimensional regularisation scheme $C_i = 5/6$ for gauge
bosons and $C_i = 3/2$ for scalars and fermions.  We also add a counterterm to
ensure that $v_H = \mu/\sqrt{\lambda}$, $m_h^2 = 2 \mu^2$ and that $m_S$ 
is given by its tree level value at $T=0$. The counterterm is
\begin{align}
  V^\text{CT}(h,S) &=
    - \frac{1}{2}\delta_\mu h^2
    + \frac{1}{4}\delta_\lambda h^4 -
      \frac{1}{2}\delta_{\mu_S} S^2,
  \label{eq:app:VCT}
\end{align}
where the factors $\delta_i$ are
\begin{align}
  \delta_\mu &=
    \frac{3}{2 v} \be{\frac{\partial V^\text{CW}}{\partial h}}{v} 
  - \frac{1}{2}\be{\frac{\partial^2 V^\text{CW}}{\partial h^2}}{v}\,, \\
  \delta_\lambda &=
    \frac{1}{2 v^3} \bigg(\be{\frac{\partial V^\text{CW}}{\partial h}}{v}
  - v \be{\frac{\partial^2 V^\text{CW}}{\partial h^2}}{v} \bigg) \,,\\
  \delta_{\mu_S} &=
    \be{\frac{\partial^2 V^\text{CW}}{\partial S^2}}{v}\,.
\end{align}
The one-loop finite temperature correction is~\cite{Dolan:1973qd}
\begin{align}
  V^T(h,S) &= \sum_i \frac{n_i T^4}{2\pi^2}\int_0^\infty \! dx \, x^2
              \log \bigg[ 1 \pm
                \exp\Big( -\sqrt{x^2+m_i^2(h,S)/T^2} \Big) \bigg] \,,
  \label{eq:app:VT}
\end{align}
where we sum over the same eigenvalues as for the Coleman--Weinberg
contribution.  The negative sign in the integrand is for bosons while the
positive sign is for fermions.

The bosons also contribute to higher order ``daisy'' diagrams which can be
resummed to give~\cite{Carrington:1991hz, Quiros:1999jp, Ahriche:2007jp,
Delaunay:2007wb}
\begin{align}
  V^\text{daisy} &= -\frac{T}{12\pi} \sum_i n_i
    \Big( \left[ m^2(h,S) + \Pi(T) \right]_i^\frac{3}{2}
    - \left[ m^2(h,S) \right]_i^\frac{3}{2}
  \Big) \,.
  \label{eq:app:Vdaisy}
\end{align}
Here, the first term should be interpreted as the $i$-th eigenvalue of the
matrix-valued quantity $[m^2(h,S) + \Pi(T)]^{3/2}$, where $m^2(h,S)$ is the
block-diagonal matrix composed of the individual mass matrices~\cite{Patel:2011th}.
The sum runs over the bosonic degrees of freedom.
The thermal (Debye) masses in the SM \cite{Carrington:1991hz} are
\begin{align}
  \Pi_{h,G^0, G^+}  &= \tfrac{1}{16} T^2 \left(3 g^2 + g'^2
                     + 8\lambda_H + 4 y_t^2\right) \,,\\
  \Pi_{W^{1,2,3}}^L &= \tfrac{11}{6} g^2 T^2 \, ,\\
  \Pi_{W^{1,2,3}}^T &= 0 \,,\\
  \Pi_B^L           &= \tfrac{11}{6} g'^2 T^2 \,, \\
  \Pi_B^T           &= 0 \,.
\end{align}
Here, $\Pi_{h, G^0, G^+}$ denotes the Debye masses of the components of $H$,
while $\Pi_{W,B}$ are the Debye masses of the electroweak gauge boson.  Note
that the latter are non-zero only for the longitudinal components of the gauge
bosons ($\Pi_{W,B}^L(T) \neq 0$), but vanish for the transverse components
($\Pi_{W,B}^T(T) = 0$).

\section{Boltzmann Equations}
\label{sec:app-boltzmann}

In the following, we discuss the Boltzmann equations governing DM freeze-in.

\subsection{Freeze-In via Decay: \boldmath$B_1 \to \chi B_2$}
\label{sec:B1-chi-B2}

For a $1 \to 2$ decay of the form $B_1 \to \chi B_2$ (where $\chi$ is the DM
candidate and $B_1$, $B_2$ are its interaction partners), the general Boltzmann
equation is
\begin{align}
  \dot{n}_\chi + 3 H n_\chi
    &= \int\!d\Pi_\chi d\Pi_{B_1} d\Pi_{B_2} (2\pi)^4 \,
       \delta^4(p_\chi - p_{B_1} - p_{B_2}) \times
                                                \notag\\ 
  &\hspace{0.5cm}
       \Big[ |\mathcal{M}_{B_1 \to B_2 \chi}|^2 f_{B_1} (1\pm f_{B_2}) (1 \pm f_\chi)
           - |\mathcal{M}_{B_2 \chi \to B_1}|^2 f_{B_2} f_\chi (1\pm f_{B_1}) \Big].
  \label{eq:boltzmann-1-to-2}
\end{align}
Here, $d\Pi_i \equiv d^3p_i / (2\pi)^3$ denotes the three-dimensional phase space
integral for particle $i = \chi,\, B_1,\, B_2$; $p_i$ are the particles' momenta
and $f_i$ their momentum distribution functions in the primordial plasma.
$\mathcal{M}_{B_1 \to B_2 \chi}$ and $\mathcal{M}_{B_2 \chi \to B_1}$ are the
transition amplitudes of the freeze-in reaction and its inverse.
In freeze-in scenarios, the abundance of $\chi$ remains well below its equilibrium 
abundance, so $f_\chi \simeq 0$. Consequently, the second term in square brackets
in \cref{eq:boltzmann-1-to-2} can be dropped. We also neglect Pauli blocking and 
stimulated emission, so that $(1 \pm f_{B_i}) = 1$. We can then write 
\begin{align}
  \dot{n}_\chi + 3 H n_\chi
    &= \int\! d\Pi_{B_1} g_{B_1} \,
       2 m_{B_1} \Gamma(B_1 \rightarrow B_2 \chi) \, f_{B_1} \,.
\end{align}
We assume that $B_1$ is in thermal equilibrium and that $E_{B_1} \gg T$ so 
that $f_{B_1} \simeq e^{-E_{B_1}/T}$.  We expect this assumption to introduce an 
error of around 15\% in our final abundances~\cite{Blennow:2013jba}.
We obtain
\begin{align} 
  \dot{n}_\chi + 3Hn_\chi
    &= \frac{g_{B_1} m_{B_1}^2}{2\pi^2}
       \Gamma(B_1\rightarrow \chi B_2) \, T \,
       K_1\! \Big(\frac{m_{B_1}x}{m_\chi}\Big) \,,
\end{align}
where $K_1$ is a modified Bessel function of the second kind
and $g_{B_1}$ is the number of degrees of freedom of $B_1$.
It is convenient to normalise the number densities $n_i$ to the entropy
density $s$ to factorise the trivial dilution of $n_i$ due to the expansion
of the Universe.  This leads to the yield $Y_i \equiv n_i / s$.
Introducing the dimensionless evolution variable $x \equiv m_\chi /T$,
the Boltzmann equation takes its final form
\begin{align} 
  \frac{d Y_\chi}{d x}
    &= \frac{g_{B_1} m_{B_1}^2}{2\pi^2} \frac{m_\chi}{H(x) s(x) x^2} 
       \Gamma(B_1\to \chi B_2) \, K_1\!\Big(\frac{m_{B_1}x}{m_\chi}\Big) \,,
  \label{eq:BM1}
\end{align}
where $H(x)$ is the Hubble rate.

\subsection{Freeze-In via Annihilation: \boldmath$B_1 B_2 \to \chi B_3$}
\label{sec:B1-B2-chi-B3}

Following similar steps as in \cref{sec:B1-chi-B2}, and using the definition of 
the M\o ller velocity $v_\text{M\o l}$~\cite{Gondolo:1990dk}, the Boltzmann
equation for a $2 \to 2$ process of the form $B_1 B_2 \to \chi B_3$ reads
\begin{align}
  \dot{n}_\chi + 3 H n_\chi
    &= g_{B_1} g_{B_2} \int\!\frac{dp^3_{B_1} dp^3_{B_2}}{(2\pi)^6}
       \sigma(B_1 B_2 \to \chi B_3) \, v_\text{M\o l} e^{-(E_{B_1} + E_{B_2})/T} \,.
  \label{eq:boltzmann-2-to-2}
\end{align}
We simplify this expression following ref.~\cite{Edsjo:1997bg}. To this
end, we define
\begin{align}
  E_+ &\equiv E_{B_1} + E_{B_2}\,,\\ 
  E_- &\equiv E_{B_1} - E_{B_2}\,,\\ 
  s   &\equiv m_{B_1}^2 + m_{B_2}^2 + 2 E_{B_1} E_{B_2}
                 - 2 |\vec{p}_{B_1}| \, |\vec{p}_{B_2}| \cos\theta\,,
\end{align}
where $E_i$ and $\vec{p}_i$ are the particles' energies and three-momenta, respectively,
and $\theta$ is the angle between $\vec{p}_{B_1}$ and $\vec{p}_{B_2}$.
It is straightforward to show that
\begin{align}
  d^3p_{B_1} d^3p_{B_2} &= 2 \pi^2 E_{B_1} E_{B_2} dE_+ dE_- ds \,.
\end{align}
Moreover, we have~\cite{Edsjo:1997bg}
\begin{align}
  E_{B_1} E_{B_2} \sigma(B_1 B_2 \to \chi B_3) \, v_\text{M\o l}
    &= \sigma(B_1 B_2 \to \chi B_3) \, p_{B_1 B_2} \sqrt{s} \,,
\end{align}
where
\begin{align}
  p_{ij} &\equiv \frac{\sqrt{[s - (m_i+m_j)^2] [s - (m_i-m_j)^2]}}{2\sqrt{s}}\,
\end{align}
is the modulus of the momentum of $B_1$ and $B_2$ in the centre of mass frame.
The integrand on the right hand side of the Boltzmann equation
\eqref{eq:boltzmann-2-to-2} is independent of $E_-$ and depends on
$E_+$ only through the exponential $e^{-E_+/T}$. Evaluating the integrals
over $E_+$ and $E_-$ yields~\cite{Edsjo:1997bg}
\begin{align}
  \dot{n}_\chi + 3 H n_\chi
    &= g_{B_1} g_{B_2} \int_{(m_{B_1} + m_{B_2})^2}^\infty
       \frac{ds}{32 \pi^4}  4 p_{B_1 B_2}^2
       \sigma(B_1 B_2 \to \chi B_3) T K_1\bigg(\frac{\sqrt{s}}{T}\bigg)\,.
  \label{eq:boltzman-2-to-2-result-1}
\end{align}
Note that, in counting the degrees of freedom $g_{B_1}$, $g_{B_2}$, care must
be taken that each degree of freedom counted towards $g_{B_1}$ should be
able to annihilate with each degree of freedom counted towards $g_{B_2}$.
This is usually true for spin and colour degrees of freedom (which the cross
sections are typically averaged over). Particles and antiparticles, on the
other hand, should be treated as different initial states, not as different
degrees of freedom of the same initial state. The same is true for different
components of an $su(2)_L$ multiplet.

Expressed in terms of particle yields rather than number densities,
\cref{eq:boltzman-2-to-2-result-1} transforms into
\begin{align} 
  \frac{d Y_\chi}{d x}
    &= \frac{g_{B_1} g_{B_2}}{32\pi^4} \frac{m_\chi}{H(x) s(x) x^2} \,
       \int_{(m_{B_1} + m_{B_2})^2}^\infty \! ds \,
       4 p_{B_1 B_2}^2  \sigma(B_1 B_2 \to \chi B_3) \,
       K_1\bigg(\frac{x\sqrt{s}}{m_\chi}\bigg) \,.
\end{align}
For the special case $B_2 = \bar{B}_1$, this simplifies to
\begin{align} 
  \frac{d Y_\chi}{d x}
    &= \frac{g_{B_1}^2}{32\pi^4} \frac{m_\chi}{H(x) s(x) x^2}
       \int_{4 m_{B_1}^2}^\infty \! ds \,
       (s - 4 m_{B_1}^2) \sigma(B_1 \bar{B}_1 \to \chi B_3) \,
       \sqrt{s} \, K_1\bigg(\frac{x\sqrt{s}}{m_\chi}\bigg) \,.
\end{align}

\bibliography{./refs}

\end{document}